\documentclass[onecolumn]{aastex61}
\submitjournal{ApJ}

\begin{document}

\title{Epicyclic oscillations in the Hartle--Thorne external geometry}

\author{Gabriela  Urbancov\'{a}}
\affiliation{Institute of Physics, Faculty of Philosophy and Science, Silesian University in 
Opava, Bezru\v{c}ovo n\'{a}m. 13, CZ-74601 Opava, CZ}
\affiliation{Astronomical Institute of the Czech Academy of Sciences, Bo\v{c}n\'{\i} II 1401, 
CZ-14100 Prague, CZ}

\author{Martin Urbanec}
\affiliation{Institute of Physics, Faculty of Philosophy and Science, Silesian University in 
Opava, Bezru\v{c}ovo n\'{a}m. 13, CZ-74601 Opava, CZ}

\author{Gabriel T\"{o}r\"{o}k}
\affiliation{Institute of Physics, Faculty of Philosophy and Science, Silesian University in 
Opava, Bezru\v{c}ovo n\'{a}m. 13, CZ-74601 Opava, CZ}

\author{Zden\v{e}k Stuchl\'{\i}k}
\affiliation{Institute of Physics, Faculty of Philosophy and Science, Silesian University in 
Opava, Bezru\v{c}ovo n\'{a}m. 13, CZ-74601 Opava, CZ}

\author{Martin Blaschke}
\affiliation{Institute of Physics, Faculty of Philosophy and Science, Silesian University in 
Opava, Bezru\v{c}ovo n\'{a}m. 13, CZ-74601 Opava, CZ}

\author{John C. Miller}
\affiliation{Department of Physics (Astrophysics), University of Oxford, Keble Road, Oxford 
OX1 3RH, UK}

\correspondingauthor{Gabriela Urbancov\'{a}}
\email{gabriela.urbancova@fpf.slu.cz}

\begin{abstract}
  
The external Hartle--Thorne geometry, which describes the space-time outside a 
slowly-rotating compact star, is characterized by the gravitational mass $M$, angular 
momentum $J$ and quadrupole moment $Q$ of the star and gives a convenient description which, 
for the rotation frequencies of more than 95\% of known pulsars, is sufficiently accurate for 
most purposes. We focus here on the motion of particles in these space-times, presenting a 
detailed systematic analysis of the frequency properties of radial and vertical epicyclic 
motion and of orbital motion. Our investigation is motivated by X-ray observations of binary 
systems containing a rotating neutron star which is accreting matter from its binary 
companion. In these systems, twin high-frequency quasi-periodic oscillations are sometimes 
observed with a frequency ratio approaching $3:2$ or $5:4$ and these may be explained by 
models involving the orbital and epicyclic frequencies of quasi-circular geodesic motion. In 
our analysis, we use realistic equations of state for the stellar matter and proceed in a 
self-consistent way, following the Hartle--Thorne approach in calculating both the 
corresponding values of $Q$, $M$ and $J$ for the stellar model and the properties of the 
surrounding spacetime. Our results are then applied to a range of geodetical models for QPOs.

A key feature of our study is that it implements the recently-discovered universal relations 
among neutron star parameters so that the results can be directly used for models with 
different masses $M$, radii $R$ and rotational frequencies $f_\mathrm{rot}$.
 
\end{abstract}

\keywords{Hartle--Thorne model --- QPOs --- stars: epicyclic oscillations --- 
stars: pulsars --- X-rays: binaries}

\section{Introduction}
\label{intro}
Exterior space-times of rotating compact stars have been studied extensively for many years. 
Pioneering work was done by Hartle and Thorne who developed a slow-rotation approximation 
\citep{Har:1967:ApJ:,Har-Tho:1968:ApJ:}, describing the structure of the star itself and also 
of the surrounding vacuum space-time, constructed as a perturbation of a corresponding 
spherically-symmetric non-rotating solution, with the perturbation being taken up to second 
order in the star's angular velocity $\Omega$. In this approximation the exterior space-time 
is fully determined by the gravitational mass $M$, angular momentum $J$ and quadrupole moment 
$Q$ of the rotating star, with the inner boundary of the exterior region being given by the 
stellar radius $R$. 

Solutions for the equivalent problem without the restriction to slow rotation can be 
constructed using numerical relativity codes such as RNS (\citet{RNS}) and LORENE/nrotstar 
(\citet{Lorene1,Lorene}), but there are certain advantages to working with the Hartle--Thorne 
method, including that of having an analytic solution for the vacuum metric outside the star. 
The main topic of this paper concerns the motion of particles around rotating neutron stars, 
and we use the rotationally induced change in the frequency with which a test particle orbits 
the star at the marginally stable circular orbit as a key quantity for comparing different 
approaches. Comparison of the Hartle--Thorne values for this with those coming from the 
numerical relativity codes shows good agreement for dimensionless spins $j=J/M^2 \simeq 
0.5$\footnote{Throughout this paper, we are using units for which $c=G=1$.}, suggesting 
applicability of other results obtained using the Hartle--Thorne approximation in most 
astrophysically relevant situations, although this is something which requires further 
checking. Comparisons between numerical-relativity and analytical space-times have been 
discussed by a number of authors (see, for example, \citet{Noz-Ste-Gou:1998:AAPS:, 
Ber-Ste:MNRAS:2004:,Ber-etal:MNRAS:2005:}). Something not considered in the present paper is 
the role of the magnetic field of a neutron star for affecting the motion of particles in its 
vicinity. Discussion of this can be found in articles by \citet{San-etal:2010:, 
Bak-etal:CQG:2012:,Gut-Val-Pac:2013:} and others.

Details and overviews of rotating neutron stars and related physics can be found especially 
in the books by \citet{Web:PSRBOOK:} and \citet{Fri-Ste:BOOK:} and in the review article by 
\citet{Pasch-Ster:2017:LRR:}.

Rotating neutron stars are usually observed as pulsars, either isolated or in binary systems. 
In binary systems they can reach high rotation frequencies due to accretion of matter from 
the binary companion which increases both the star's gravitational mass and its angular 
momentum (and hence its rotational speed). The rotational frequencies can reach hundreds of 
Hz (the fastest currently known pulsar rotates with frequency 716Hz 
\citep{Hes-etal:2006:Science:}). Within this work we are interested mostly in neutron stars 
which are in Low Mass X-ray Binaries, where the neutron star is accompanied by a low-mass 
ordinary star. Observations of some of these objects exhibit the phenomenon of twin 
high-frequency quasi-periodic oscillations (QPOs) \citep{Kli:2006:CompStelX-Ray:}.

These QPOs consist of two adjacent peaks in the X-ray power spectrum and have been observed 
in a number of sources. They often have a lower frequency of $600-800\,\mathrm{Hz}$ and an 
upper frequency of $900-1200\,\mathrm{Hz}$, and there is clustering of the observed frequency 
ratios particularly at around $3:2$ but also at around $4:3$ and $5:4$ 
\citep{Bel-etal:2007:MONNR:RossiXTE,Tor-Bak-Stu-Cec:2008:ACTA:TwPk4U1636-53:,
Tor-etal:2008:ACTA:DistrKhZ4U1636-53,Tor-etal:2008:ACTA:Clustering4U1636-53,
Bou-etal:2010:MONNR}. Similar clustering is observed in microquasars (low mass X-ray binaries 
containing black holes) where the frequencies are fixed at the ratio of $3:2$ and can be 
explained as resulting from a non-linear resonance between radial and vertical oscillation 
modes of an accretion disc around the black hole \citep{Tor-Abr-Klu-Stu:2005:ASTRA:}. 
Therefore, we can anticipate that the $3:2$ ($4:3$ and $5:4$) resonance should play a role 
also in systems containing neutron stars instead of black holes 
\citep{Tor-etal:2008:ACTA:Clustering4U1636-53}. Nevertheless, it is still unclear whether or 
not the clustering in the distribution of frequency ratios is a real effect 
\citep{gilf-etal:2003,men:2006,Bou-etal:2010:MONNR,Mon-Zan:2012:MONRAS:,Ribeiro-etal:2017}.

There have been several attempts to explain the physical origin of QPOs using simple orbital 
models based on motion of matter around a central object (e.g.\,\citet{Ste-Vie:1998:ApJ:,
Ste-Vie:1999:PHYSRL:,wag:1999,ste-etal:2001,abr-klu:2001,klu-abr:2001b,wag-etal:2001,
abr-etal:2003b}. A large set of references to models for neutron-star QPOs, along with a 
direct comparison between them, can be found in \citet{Tor-etal:2010:ApJ:,
Tor-etal:2012:ApJ:,Tor-etal:2016:MNRAS:}). A more elaborate model which included corrections 
to the orbital motion due to finite thickness of the accretion disk has been investigated by 
\citet{Tor-etal:2016:MNRAS:} and was particularly successful in the case of 4U1636-53 where 
it was shown that the observed frequency pairs could be explained by a torus with its center 
(maximum of density) located at positions which were variable but always at large enough 
radius to allow accretion onto the central object via the inner cusp; this model has been 
called the cusp torus model (see the papers by \citet{rez-etal:2003,abr-etal:2006,sra:2007,
Ingram+Done:2010,fragile-blaes:2016,Tor-etal:2016:MNRAS:,mis-etal:2017,
Tor-etal:2017a:arXiv:,Tor-etal:2017b:RAGtime:,Parth-etal:2017:,ave-etal:2017} and references 
therein). Within this model the highest observed frequencies correspond to the torus being 
located close to the marginally stable orbit and being small enough so that the correction 
due to the finite thickness of the disk could be neglected, enabling one to put constraints 
on the radius of the central object.

Motivated by the cusp torus model, we here make a detailed study of quasi-circular epicyclic 
motion in the external Hartle--Thorne geometry, using the formulas obtained previously by 
\citet{Alm-Abr-Klu-Tam:2003:ArXiv:}. We begin our analysis by using the Hartle--Thorne 
approach to calculate neutron-star properties for a range of realistic equations of state for 
the neutron-star matter, and demonstrate how the influence of the different prescriptions for 
the equation of state can be hidden if one plots particular combinations of neutron star 
properties against one another. These results are manifestations of the universal relations 
which have been discovered rather recently \citep{YY1,YY2,Urb-Mil-Stu:2013:MONNR:,
Mas-etal:2013:PRD:,Pap:2015:MNRAS:,Rei-etal:2017:MNRAS:} and have been studied extensively 
under various circumstances \citep{Cha-etal:2014:PRL:,Has-etal:2014:MNRAS:,
Don-etal:2014:PRD:,Don-etal:2014:APJL:,Sha-etal:2015:APJ:,Don-Yas-Kok:2015:PRD:,
Pan-Gua-Fer:2015:PRD:,Sil-Sot-Ber:2016:MNRAS:,Sta-Don-Yaz:2016:PRD:}.

We then move on to discuss properties of particle motion around the star, showing the radial 
profiles for the orbital (Keplerian) frequency of circular motion and the radial and vertical 
frequencies of epicyclic quasi-circular motion. These profiles are relevant down to the 
radius of the innermost stable circular geodesic. We then study the frequency ratio of the 
oscillatory modes related to various geodetical models which have been proposed for 
explaining twin HF QPOs: the relativistic precession model and its variants 
\citep{Ste-Vie:1998:ApJ:,Ste-Vie:1999:PHYSRL:}, an epicyclic resonance model 
\citep{Tor-Abr-Klu-Stu:2005:ASTRA:}, a tidal disruption model 
\citep{Kos-Cad-Cal-Gom:2009:ASTRA} and a warped disc oscillation model 
\citep{Kat:2004:PASJ:}\footnote{The epicyclic frequencies in the case of fully numerical 
models for rapidly rotating neutron stars have been studied in \citep{Pap:2012:MONNR:} and 
\citep{Gon-Klu-Ste-Wis:2014:PHYSR4:}.}.

In Section\,\ref{sec:1} we give a brief introduction to the Hartle--Thorne methodology and 
then present results for neutron-star models constructed using our set of equations of state. 
In Section\,3, we present general properties of the radial profiles for the orbital 
frequency, the epicyclic frequencies and the precession frequencies and also describe the 
behaviour of the maximum of the profile for the radial epicyclic frequency. In Section\,4 we 
show radial profiles for the frequency ratios corresponding to various models for HF QPOs and 
in Section\,5 we compare models calculated using the Hartle--Thorne approach with fully 
numerical models not restricted to slow rotation. The main results presented within Sections 
1-5 are parametrized in terms of the NS mass, angular momentum and quadrupole moment. This 
allows their convenient application for interpreting neutron-star X-ray variability data 
within the context of the universal relations. We end with a short Summary in Section\,6.

\section{The Hartle--Thorne approach for calculating models of rotating compact stars}

\label{sec:1}
The Hartle--Thorne approach \citep{Har:1967:ApJ:,Har-Tho:1968:ApJ:} is a convenient scheme 
for calculating models of compact stars in slow and rigid rotation and, since its 
introduction, has been used and discussed by many other authors. In this scheme, deviations 
away from spherical symmetry are considered as being small perturbations and treated by means 
of series expansions with terms up to second order in the star's angular velocity $\Omega$ 
being retained.

The metric line element takes the form (in standard Schwarzschild coordinates):

\begin{eqnarray}
\mathrm d s^2 &=& -\mathrm e^{2\nu}[1+2h_0(r)+2h_2(r)P_2] \mathrm dt^2 +
\mathrm e^{2\lambda}\left\{1+\frac{\mathrm e^{2\lambda}}{r}[2m_0(r)+2m_2(r)P_2]\right\}
\mathrm d r^2 \nonumber \\
&&+r^2\left[1+2k_2(r)P_2\right]\left\{\mathrm d \theta^2 + 
[\mathrm d \phi-\omega(r)\mathrm d t]^2\sin^2 \theta \right\},
\label{eq:HTMetricInternal}
\end{eqnarray}
 where $\nu$ and $\lambda$ are functions of the radial coordinate $r$, and reduce to those of 
the spherically symmetric vacuum Schwarzschild solution in the exterior, $\omega(r)$ is a 
perturbation of order $\Omega$ which represents the dragging of inertial frames, and the 
perturbation functions $h_{0}(r)$, $h_{2}(r)$, $m_{0}(r)$, $m_{2}(r)$ and $k_{2}(r)$ are of 
order $\Omega^2$ and are functions only of the radial coordinate with the non-spherical 
angular dependence on the latitude $\theta$ being given by the Legendre polynomial $P_{2} = 
P_{2}(\theta) = [3 \cos^2(\theta) - 1]/{2}$.

The interior metric is found by solving the Einstein equations with the stress-energy tensor 
on the right hand side being that for a perfect fluid rotating rigidly with angular velocity 
$\Omega$, while the exterior solution is found by solving the vacuum Einstein equations, and 
constants appearing in the calculations are determined by matching these two solutions at the 
surface of the star.\footnote{As pointed out quite recently by \citet{Rei-Ver:2015:CQG:} and 
\citet{Rei:2016:MNRAS:}, special care is required when doing this for models having a density 
discontinuity at the surface in order to avoid getting wrong values for rotational 
corrections to the mass, but this issue does not arise for the models being discussed here.} 
The exterior solution can be expressed in terms of the star's gravitational mass $M$, angular 
momentum $J$ and quadrupole moment $Q$ \footnote{Note that throughout we follow the convention of taking $Q$ to be {\em positive for an oblate object}, as used by Hartle \& Thorne.} 
and it can be useful to rewrite the external metric in the terms of the 
dimensionless quantities: $j=J/M^2$, a dimensionless angular momentum, and $q=Q/M^3$, a 
dimensionless quadrupole moment \citep{Alm-Abr-Klu-Tam:2003:ArXiv:}. Quadrupole moments of 
rotating stars have been under investigation in the past and various definitions can be 
applied, for more details see e.g. the discussions by \citet{Bon-Gou:1996:AAP:} and 
\citet{Pap-Apo:PRL:2012:,Fri-Ste:BOOK:}.

Within our calculations we also introduce another quadrupole parameter $\tilde q = q/j^2 = 
QM/J^2$ which is also dimensionless. Since $q \sim \Omega^2$ and $j \sim \Omega$, 
$\tilde q$ is independent of the star's rotation speed within the Hartle--Thorne 
approximation. (A similar property has been shown to apply for rapidly rotating neutron 
stars as well, extending also to appropriately scaled higher order moments 
\citep{Pap-Apo:PRL:2012:,Pap-Apo:ArXiv:2012:,Pap-Apo:PhRvL:2014,
Yagi-Kyu-Pap-You-Apo:PhRvD:2014:}.)

\subsection{Models of rotating neutron stars}

Here we present results of calculations made using the Hartle--Thorne approach in order to 
determine the relations satisfied by the main neutron-star properties for a range of 
equations of state. For each EoS we show a sequence of models rotating at 400Hz (with masses 
up to the maximal one) and a second similar sequence with zero rotation, for comparison. 
Models satisfying the slow-rotation criterion will lie between the two (the motivation for 
focusing on 400 Hz will be explained later). The models of rotating neutron stars are 
constructed by solving a set of ordinary differential equations for the perturbation 
functions; the approach used has been extensively described in the literature and we will not 
repeat that here, see e.g. \citet{Har-Tho:1968:ApJ:,Chan-Mil:1974:MONNR:,Mil:1977:MONNR:,
Web-Gle:1992:APJ:,Ben-etal:PRD:2005:,Pasch-Ster:2017:LRR:}.

\begin{figure}
  \includegraphics[width=0.46\textwidth]{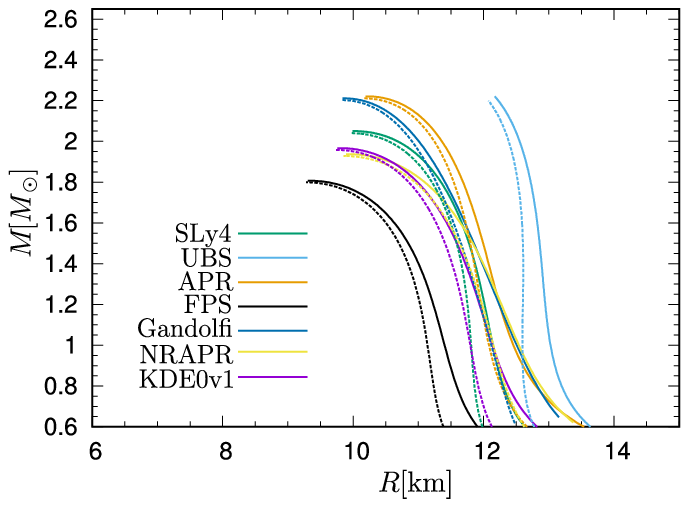}
  \includegraphics[width=0.49\textwidth]{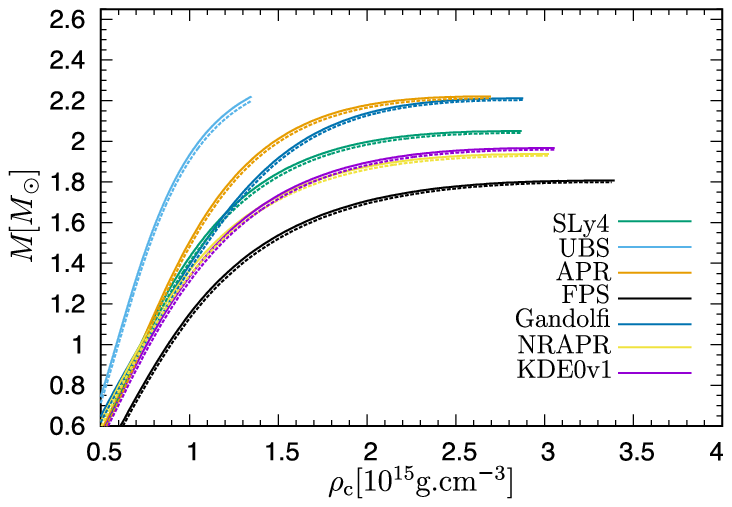}

  \includegraphics[width=0.47\textwidth]{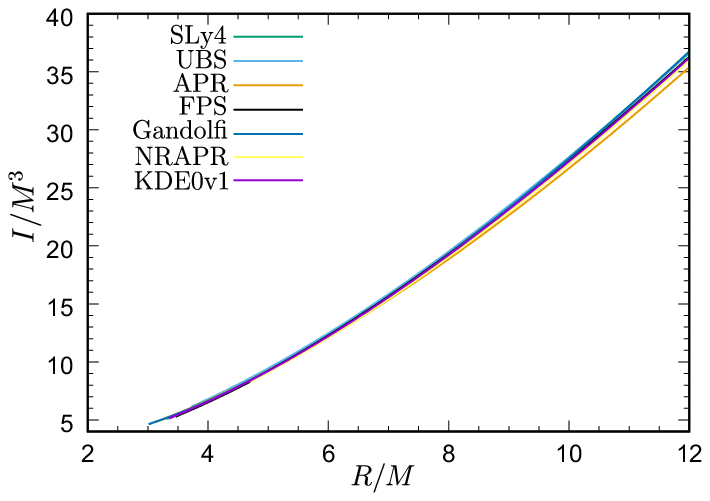}
  \includegraphics[width=0.47\textwidth]{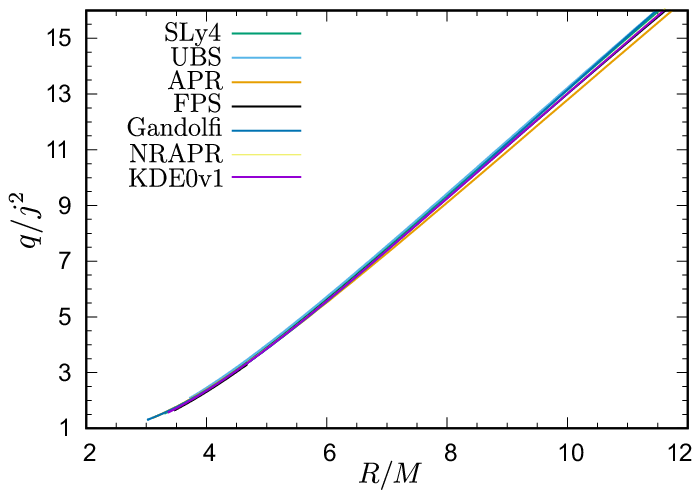}

  \caption{{\it Top:} Mass versus radius (left) and mass versus central energy density 
(right) are plotted for a selection of equations of state, with the sequences of rotating and 
non-rotating stars being represented with solid and dashed lines respectively. The rotating 
ones all have a rotation frequency of 400Hz and the value of $R$ given for them is the 
equatorial radius ($R = R_\mathrm{eq}$). {\it Bottom:} The dimensionless quantities $I/M^3$ 
(left) and $q/j^2 = QM/J^2$ (right) are plotted against the inverse compactness $R/M$. All of 
the quantities in the bottom panels are in geometrical units, with $c=G=1$.}
  \label{fig:statmodels1}
\end{figure}

For our set of equations of state describing nuclear matter we have selected SLy4 
\citep{Rik-Mil-Kon-Ste-Str:2003:PHYSRC:}, UBS \citep{Urb-Bet-Stu:2010:ACTA:}, APR 
\citep{APR}, FPS \citep{FPS}, Gandolfi \citep{Gan-Illa-Fan-Mil-Ped-Schm:2010:MONNR:}, NRAPR 
\citep{NRAPR} and KDE0v1 \citep{KDE}. These follow various approaches to nuclear matter 
theory and lead to neutron star models with different properties. Fig.\ref{fig:statmodels1} 
shows mass-radius relations (top left) and mass as a function of the central energy density 
(top right) for both rotating models (solid lines) and non-rotating models (dashed lines). 
The different colours correspond to the different equations of state. For the rotating 
models, the equatorial radius is the quantity plotted as $R$. By comparing the right and left 
panels, one can see that the more compact stars (having smaller total radius for a given 
mass) are also having higher central energy densities, as would be expected. From the right 
panel one can see the mass becoming less sensitive to central density as the maximum is 
approached. The existence of the maximum mass is important for testing equations of state 
using astronomical observations: quite recent observations of two solar mass neutron stars 
\citep{demorest,antoniadis} rule out many equations of state and from our selection of 
nuclear matter models one can see that FPS is not applicable for these two objects.

The bottom panels of Fig.\ref{fig:statmodels1} show the extent to which particular 
combinations of neutron star parameters are universally related to the inverse compactness 
$R/M$ (we are again using geometrical units here, so that all quantities are measured 
in units of length or its powers). As we will see, making use of this quasi-universal 
behaviour can be extremely convenient for investigating orbital motion. For given $R$ and $M$, 
one can immediately obtain the value of $I/M^3$ from the relation in the bottom left panel of 
Fig.\ref{fig:statmodels1} and hence calculate the dimensionless angular momentum $j=J/M^2= 
2\pi f_\mathrm{rot}I/M^2$ for particular values of star's rotational frequency 
$f_\mathrm{rot}$. The dimensionless quadrupole moment $q=Q/M^3$ can then be obtained by using 
the relation in the bottom right panel together with the value of $j$.

This approach enables one to skip making the calculations of neutron-star models for all 
possible equations of state and central parameters, and to use just global properties of the 
star (its mass and radius) for calculating all parameters of the external spacetime for the 
required rotational frequency.

\begin{figure}
  \includegraphics[width=0.49\textwidth]{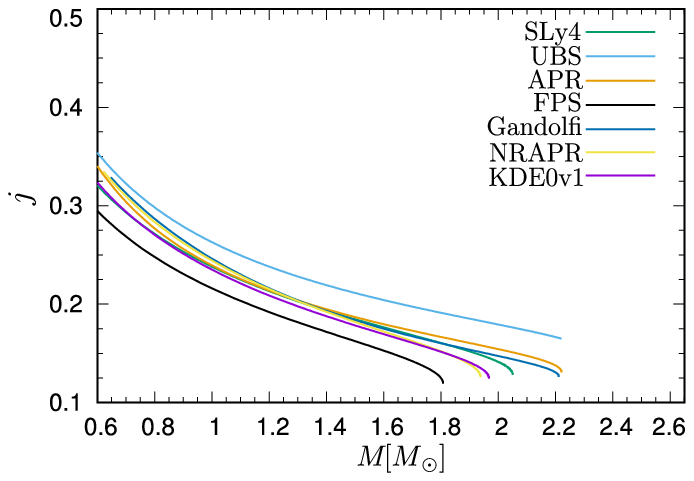}
  \includegraphics[width=0.49\textwidth]{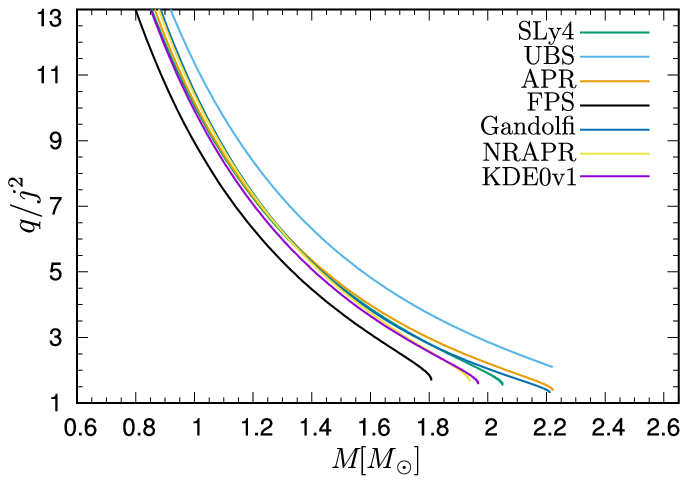}

  \caption{Values of $j=J/M^2$ (left) and $\tilde q = q/j^2=QM/J^2$ (right) plotted against 
gravitational mass for stars rotating at 400~Hz.}

  \label{fig:statmodels2}
\end{figure}

For illustration, in Fig.\,\ref{fig:statmodels2} we demonstrate results from detailed 
modelling: the left panel shows values of $j$ as a function of the gravitational mass for 
stars rotating with $f_\mathrm{rot}=400$Hz. Within the Hartle--Thorne approximation $j=J/M^2$ 
depends linearly on rotational frequency, and so one can easily find $j$ for other values of 
$f_\mathrm{rot}$ by simple linear scaling. The maximal value of $j$ is that for a star 
rotating at its mass-shedding limit and, for neutron stars, reaches values of 0.65--0.7 
\citep{Lo-Lin:2011:ApJ:}. The right panel shows the values of $\tilde q = QM/J^2$ as a 
function of gravitational mass. It can be seen that for neutron stars with $M \geq 
1.2\,M_\odot$, the values of $\tilde q$ are less than 10, and we will use this later as a 
maximum in our investigation of the motion of particles in the field of rotating neutron 
stars, to keep us within the limits of astrophysicaly interesting objects.

In this section we have discussed the universal behaviour of the parameters $j$ and $\tilde 
q$ and how they are related to the mass, radius and rotational frequency of neutron star 
models calculated using realistic equations of state. In the following sections, we will make 
use of this for studying the motion of particles in the space-time outside rotating neutron 
stars within the Hartle--Thorne approximation. A similar investigation, but using a different 
model for the space-time around the rotating neutron stars, has been performed by 
\citet{Pach-Rue-Valen:ApJ:2012:} who presented their results in a more condensed form than 
being done here. Our approach is appropriate for any model of the rotating neutron stars 
since we plot the results in a way that is not dependent on any particular choice of equation 
of state.

\section{The external Hartle--Thorne geometry and the frequencies of epicyclic motion}

Since our main interest here is to investigate free-particle motion external to the surface 
of rotating neutron stars, we will next discuss this part of the space-time (for $r > R$ to 
lowest order) in more detail. We should stress that all of the discussion of this section and 
the next one, apply {\em only to the region outside the surface of the neutron star}. Any 
reference to free particle motion which would seem to occur inside the neutron star needs to 
be disregarded.

As we have been discussing, the external Hartle--Thorne geometry can be expressed in terms of 
three parameters, for which we here take the gravitational mass $M$, the dimensionless 
angular momentum $j$ and the dimensionless quadrupole moment $q$, in the form 
\citep{Alm-Abr-Klu-Tam:2003:ArXiv:}\footnote{Note misprints in the original paper.}:

\begin{eqnarray}
g_{tt}&=&-(1-2M/r)[1+j^{2}F_{1}(r)+q F_{2}(r)]\nonumber, \\
g_{rr}&=&(1-2M/r)^{-1}[1+j^2G_{1}(r)-q F_{2}(r)]\nonumber, \\
g_{\theta\theta}&=&r^2[1+j^2H_{1}(r)+q H_{2}(r)]\nonumber, \\
g_{\phi\phi}&=&r^2\sin^2\theta[1+j^2H_{1}(r)+q H_{2}(r)]\nonumber,\\
g_{t\phi}&=&g_{\phi t} =-2(M^2/r)j\sin^2\theta,
\end{eqnarray}
where
\begin{eqnarray}
F_{1}(r)&=&-[8Mr^4(r-2M)]^{-1}[u^2(48M^6-8M^5r-24M^4r^2 -30M^3r^3-60M^2r^4+135Mr^5-45r^6) 
\nonumber \\
         &&+(r-M)(16M^5+8M^4r-10M^2r^3 -30Mr^4+15r^5)]+A_{1}(r),\nonumber\\
F_{2}(r)&=&[8Mr(r-2M)]^{-1}(5(3u^2-1)(r-M)(2M^2+6Mr-3r^2))-A_{1}(r),\nonumber\\
G_{1}(r)&=&[8Mr^4(r-2M)]^{-1} ((L(r)-72M^5r)-3u^2(L(r)-56M^5r))-A_{1}(r),\nonumber\\
L(r)&=&(80M^6+8M^4r^2+10M^3r^3+20M^2r^4-45Mr^5+15r^6),\nonumber\\
A_{1}(r)&=&\frac{15r(r-2M)(1-3u^2)}{16M^2}\ln\left(\frac{r}{r-2M}\right),\nonumber\\
H_{1}(r)&=&(8Mr^4)^{-1}(1-3u^2)(16M^5+8M^4r -10M^2r^3+15Mr^4+15r^5)+A_{2}(r),\nonumber\\
H_{2}(r)&=&(8Mr)^{-1}(5(1-3u^2)(2M^2-3Mr-3r^2))-A_{2}(r),\nonumber\\
A_{2}(r)&=&\frac{15(r^2-2M^2)(3u^2-1)}{16M^2}\ln\left(\frac{r}{r-2M}\right),\nonumber\\
u&=&\cos\theta.\nonumber
\end{eqnarray}

For $j=0$ and $q=0$ the external Hartle--Thorne geometry reduces to the standard 
Schwarzschild one. The Kerr geometry taken up to second order in the dimensionless angular 
momentum $a=Mj$, and using the standard Boyer-Lindquist coordinates, can be obtained from the 
external Hartle--Thorne geometry by setting $q=j^2$ and making the coordinate transformations
\begin{eqnarray}
r_\mathrm{BL} &=& r - \frac{a^2}{2r^3}[(r+2M)(r-M)+ \cos^2\theta(r-2M)(r+3M)],\\
\theta_\mathrm{BL} &=& \theta - \frac{a^2}{2r^3}(r+2M)\cos\theta \sin\theta .
\end{eqnarray}

\subsection{Radial profiles of the orbital frequency and the epicyclic frequencies}
\label{sec:radialprofiles}

Formulas for the frequencies of circular and epicyclic motion in the external Hartle--Thorne 
geometry have been calculated in \cite{Alm-Abr-Klu-Tam:2003:ArXiv:} and used by many authors, 
see e.g. \cite{Tor-Bak-Stu-Cec:2008:ACTA:TwPk4U1636-53:,Tor-etal:2014:} and 
\cite{Bos-etal:2014:GraCos:,Bos-Rue-Muc:2015:ASR:}. Here we give the relations for the 
orbital (Keplerian) frequency and the radial and vertical epicyclic frequencies, which are 
needed for the twin HF QPO models based on geodesic quasi-circular motion and were presented 
in \cite{Tor-Bak-Stu-Cec:2008:ACTA:TwPk4U1636-53:}.

The Keplerian frequency is given by
\begin{eqnarray}
\nu_{K}(r;M,j,q) &=&\frac{c^3}{2\pi GM}\frac{M^{3/2}}{r^{3/2}} 
\left[1 - j \frac{M^{3/2}}{r^{3/2}}+j^2 E_{1}(r)+q E_{2}(r)\right],
\label{eq:Kepler}
\end{eqnarray}
where
\begin{eqnarray}
E_{1}(r)&=&\frac{48M^7-80M^6r+4M^5r^2-18M^4r^3+40M^3r^4+10M^2r^5+15Mr^6-15r^7}
{16M^2(r-2M)r^4} +\frac{15(r^3-2M^3)}{32M^3}\ln\left(\frac{r}{r-2M}\right),\nonumber\\
E_{2}(r)&=&\frac{5(6M^4-8M^3r-2M^2r^2-3Mr^3+3r^4)}{16M^2(r-2M)r}
-\frac{15(r^3-2M^3)}{32M^3}\ln\left(\frac{r}{r-2M}\right),\nonumber
\end{eqnarray}

The radial epicyclic frequency $\nu_r$ and the vertical epicyclic frequency $\nu_{\theta}$ 
are given by

\begin{eqnarray}
\nu^2_{r}(r;M,j,q)&=& \left(\frac{c^3}{2\pi GM}\right)^{2}\frac{M^3(r-6M)}{r^{4}}
[1+jC_{1}(r)-j^2C_{2}(r)-qC_{3}(r)],
\label{eq:epicrad}\\
\nu^2_{\theta}(r;M,j,q)&=&\left(\frac{c^3}{2\pi GM}\right)^{2}\frac{M^3}{r^{3}}
[1- jD_{1}(r)+j^2D_{2}(r)+qD_{3}(r)],
\label{eq:epicvert}
\end{eqnarray}
where
\begin{eqnarray}
C_{1}(r)&=&\frac{6M^{3/2}(r+2M)}{r^{3/2}(r-6M)},\nonumber\\
C_{2}(r)&=&\frac{384M^8-720M^7r-112M^6r^2-76M^5r^3-138M^4r^4-130M^3r^5+635M^2r^6-375Mr^7+60r^8}
{8M^2r^4(r-2M)(r-6M)}+B_{1}(r),\nonumber\\
C_{3}(r)&=&\frac{5(48M^5+30M^4r+26M^3r^2-127M^2r^3+75Mr^4-12r^5)}{8M^2r(r-2M)(r-6M)}
-B_{1}(r)\nonumber\\
B_{1}(r)&=&\frac{15r(r-2M)(2M^2+13Mr-4r^2)}{16M^3(r-6M)}\ln\left(\frac{r}{r-2M}\right),
\nonumber\\
D_{1}(r)&=&\frac{6M^{3/2}}{r^{3/2}},\nonumber\\
D_{2}(r)&=&\frac{48M^7-224M^6r+28M^5r^2+6M^4r^3-170M^3r^4+295M^2r^5-165Mr^6+30r^7}
{8M^2r^4(r-2M)}-B_{2}(r)\nonumber\\
D_{3}(r)&=&\frac{5(6M^4+34M^3r-59M^2r^2+33Mr^3-6r^4)}{8M^2r(r-2M)}+B_{2}(r),\nonumber\\
B_{2}(r)&=&\frac{15(2r-M)(r-2M)^2}{16M^3}\ln\left(\frac{r}{r-2M}\right).\nonumber
\end{eqnarray}

We next demonstrate the dependence of the radial profiles of the orbital and epicyclic 
frequencies on the parameters $j$ and $q$. For each of the frequency profiles, we choose the 
values of the spin parameter as $j = 0.1, 0.2, 0.3, 0.5$ so as to fully cover the range which 
might be relevant for application to astrophysically interesting models of neutron stars or 
strange stars, while staying within the limit of validity of the slow-rotation approximation 
$(j/j_\mathrm{max})^2 \ll 1$. It has been shown that $j_\mathrm{max} \sim 0.65 - 0.7$ for 
neutron stars and that it could be larger for strange (quark) stars \citep{Lo-Lin:2011:ApJ:}. 
For objects which are less compact than these, the value of $j_\mathrm{max}$ could be 
significantly larger.

Values of the quadrupole parameter $q$ are calculated taking $\tilde q = q/j^2 = 
1,2,3,4,5,10$ for each of the values of $j$. This selection of values for $j$ and $q$ covers 
the relevant situations and is obtained from modeling of the neutron stars as described in 
the previous section or, for example, in \citet{Urb-Mil-Stu:2013:MONNR:}.

\begin{figure}
  \includegraphics[width=0.98\textwidth]{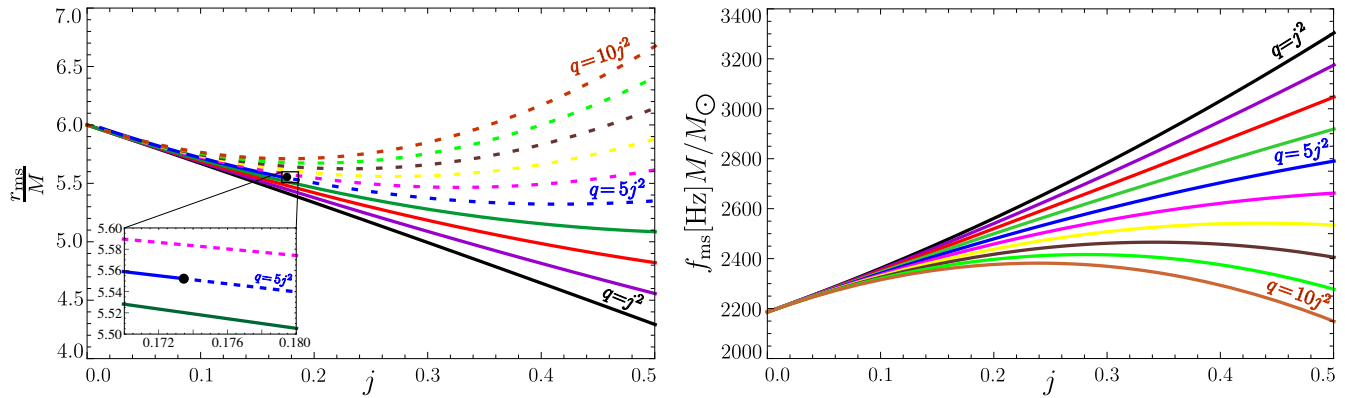}
  \caption{Left: the radial position of the marginally stable circular orbit 
$r_{\mathrm{ms}}$; dashed lines correspond to the inferred location of the marginally 
stable orbit being below the surface of the neutron star. Right: the orbital frequency 
$f_\mathrm{ms}$ at $r_\mathrm{ms}$.}
  \label{fig:Fms}
\end{figure}

In general relativity, stable circular orbital motion is restricted to radii larger than that 
of the marginally stable orbit which, in the external Hartle--Thorne geometry, is given by

\begin{eqnarray}
r_{\mathrm{ms}}=6M \left[ 1\mp j \frac{2}{3} \sqrt{\frac{2}{3}}+
j^2\left(\frac{251647}{2592}-240\ln\frac{3}{2}\right)+
q\left(-\frac{9325}{96}+240\ln\frac{3}{2}\right)\right],
\label{eq:Rms}
\end{eqnarray}

with the $-$ and $+$ referring to co-rotating and counter-rotating motion respectively. Only 
outside this orbit can quasi-circular geodesic motion be stable, giving rise to possibly 
observable quasi-periodic oscillatory effects. Properties of the marginally stable orbit have 
been discussed in detail in, for example, \citet{Tor-etal:2014:,Cip-etal:2017:PRD:} and are 
shown here in Fig.\,\ref{fig:Fms}. In the left frame, one can see that the radius of the 
marginally stable orbit $r_{ms}$ always decreases with increasing $j$ near to $j=0$, but can 
reach a minimum and then become increasing for higher values of $j$ if $q/j^2$ is large 
enough. This sort of behavior can have a relevant impact on the observed distribution of 
rotational frequencies of QPO sources \citep{Tor-etal:2014:}. The position of the 
surface of the neutron star can be calculated from the universal relations 
\citep{Yagi-Yunes:2006:ArXiv:} (given that $q/j^2$ is directly related to the inverse 
compactness of the star $R/M$); results can be seen on the left side of Fig.\,\ref{fig:Fms}. 
In the zoomed area, note the curve for $q/j^2=5$ where a dot indicates the position of the 
surface. For the rest of the specified values of $q/j^2$, $r_\mathrm{ms}$ 
is situated above the surface for $q/j^2<5$ and below the surface for $q/j^2>5$, within the 
range of $j$ shown.

The formula for the orbital frequency at $r_\mathrm{ms}$ can be calculated up to second order 
in the star's angular velocity as
\begin{eqnarray}
f_\mathrm{ms}=\frac{c^3}{2 \pi G M 6\sqrt{6}} \left[1+\frac{11j}{6\sqrt{6}}+
\frac{1}{864}j^2\left(-160583 + 397710 \ln\frac{3}{2}\right)   
+\frac{5}{32}q\left(1193 - 2946 \ln\frac{3}{2}\right) \right] .
\label{eq:fisco}
\end{eqnarray}

The right panel of Figure \ref{fig:Fms} shows $f_\mathrm{ms}$ plotted against $j$ as 
calculated from this.

In the discussion above, we have used the term ``marginally stable orbit'' thinking only of 
the external vacuum region but, in reality, for a given model of the neutron star, this 
location may not exist in the external region: the equatorial radius of the star 
$R_\mathrm{eq}$ with given parameters $M$, $j$ and $q$ can be {\em larger} than the 
$r_\mathrm{ms}$ as calculated in a vacuum exterior space-time with the same parameters. 
In the remainder of this paper, we will be using the term ``innermost stable circular 
orbit'' (ISCO) in a non standard way, to refer either to the orbit at $r_\mathrm{ms}$, if it 
exists outside the star, or otherwise to a surface-skimming orbit at $R_\mathrm{eq}$. One 
should bear in mind that, in practice, a particle on a surface-skimming orbit would be 
subject to various physical effects which we are not including here.

\begin{figure}
  \includegraphics[width=0.95\textwidth]{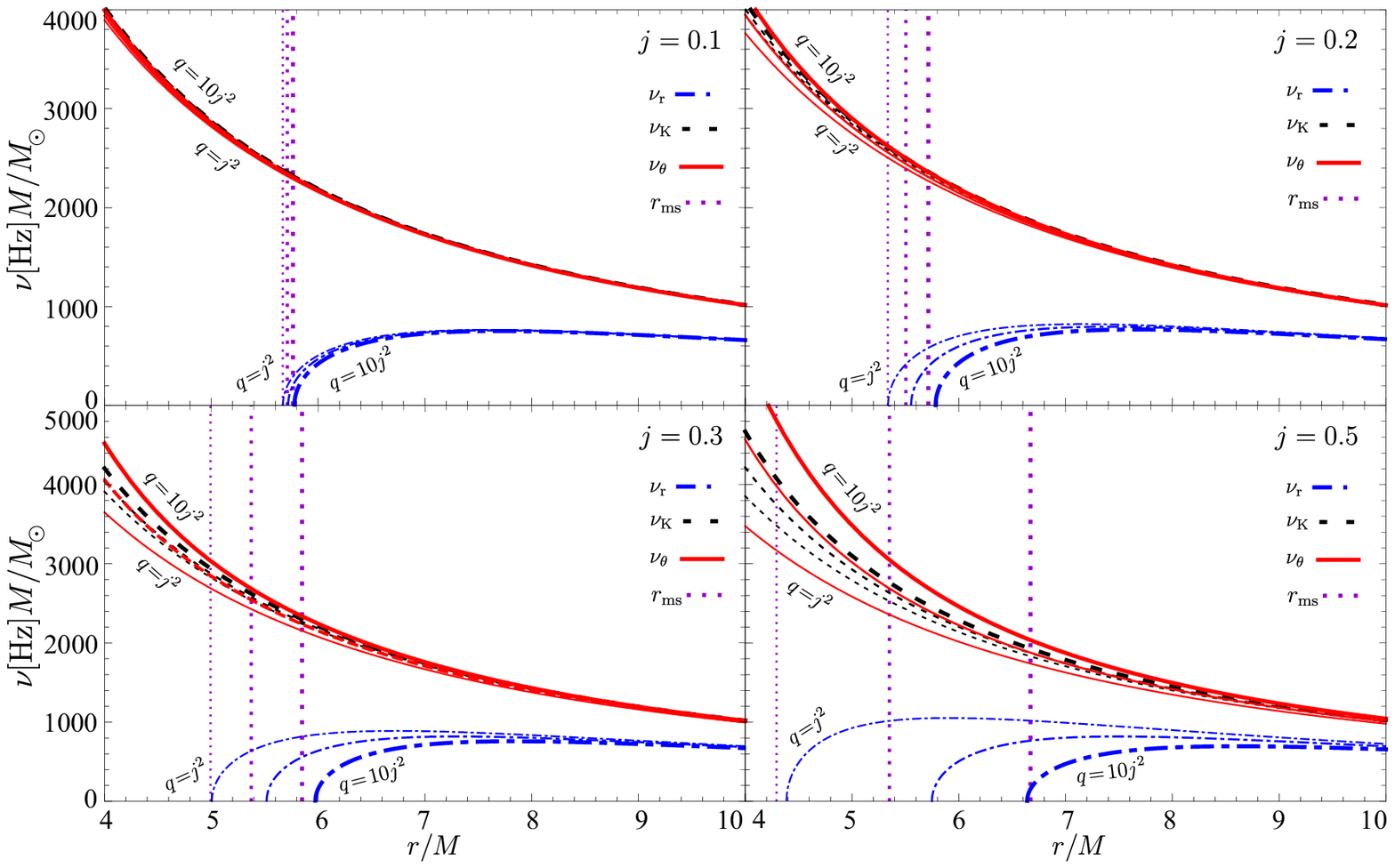}
  \caption{The radial profile of the orbital frequency $\nu_{\mathrm{K}}$ and the epicyclic 
frequencies $\nu_{\mathrm{r}}$, $\nu_{\mathrm{\theta}}$ for $j=0.1, 0.2, 0.3, 0.5$. The 
increasing thickness of the lines represents increasing values of $q/j^2= 1,~5,~10$. The 
vertical dotted lines are calculated values of $r_\mathrm{ms}$ for each $q/j^2$, as given by 
equation (\ref{eq:Rms}). Note that stable circular orbits are possible only for 
$r>r_\mathrm{ms}$. }
  \label{fig:allfreq}
\end{figure}

In order to obtain a complementary point of view, in Figure \ref{fig:allfreq} we show the 
radial profiles of the orbital and epicyclic frequencies for specified values of $j$, 
together with the profiles for the local value of the Keplerian frequency $\nu_{\mathrm{K}}$. 
One can see that the influence of the parameters $j$ and $q/j^2$ increases with decreasing 
radius, and is relatively small at radii $r \sim10M$ for all of the three frequency profiles; 
in all three cases, the influence of $q/j^2$ increases with increasing $j$. For the orbital 
frequency, the role of the parameters $j$ and $q$ is the smallest, and it is significantly 
stronger for the profile of the vertical epicyclic frequency which has a similar behavior to 
that of the orbital frequency. The most important influence is in the case of the radial 
epicyclic frequency -- the shift of the marginally stable circular orbit, corresponding to 
the radius where the radial epicyclic frequency vanishes, is significantly shifted even for 
spin $j=0.1$. For $j=0.3$ it is shifted from $r_\mathrm{ms} \sim 5M$ for $q/j^2=1$ to 
$r_\mathrm{ms} \sim 5.8M$ for $q/j^2=10$, and in the extreme case of $j=0.5$, the shift is 
from $r_\mathrm{ms} \sim 4.4M$ for $q/j^2=1$ to $r_\mathrm{ms} \sim 6.6M$ for $q/j^2=10$. 
Clearly, the role of the quadrupole moment is most significant for the radial epicyclic 
motion where it affects also the values at the maxima of radial epicyclic frequency.

One can see that, contrary to the case of the Kerr black hole spacetime, the vertical 
epicyclic frequency can be larger than the orbital frequency. This phenomenon has been 
observed also for Newtonian quadrupole gravitational fields 
\citep{Gon-Klu-Ste-Wis:2014:PHYSR4:} and general relativistic solutions with multipole 
structure of the Manko type \citep{Pap:2012:MONNR:}. A similar crossing of the radial 
profiles of the radial and vertical oscillations has been found also for axisymmetric string 
loops in the Kerr black hole spacetime \citep{Stu-Kol:2014:PHYSR4:,Stu-Kol:2015:GRG:}. This 
type of behavior was discovered by \citet{mor-ste:1999:} around rapidly rotating neutron 
stars with a very stiff equation of state and was also found for analytical approximations to 
these spacetimes \citep{Pap:2012:MONNR:,Pap-Apo:MNRAS:2013:}. Recently it has been discussed 
by \citet{Klu-Ros:2013:MNRAS:} in the context of Newtonian gravity and then further 
considered by \citet{Gon-Klu-Ste-Wis:2014:PHYSR4:} for rapidly rotating relativistic strange 
stars. Its relation to the higher order multipole moments of spacetimes around compact 
objects can be found in \cite{Pappas:MNRAS:2017:}).

Clearly, the quadrupole parameter $q/j^2$ is playing an important role and this needs to be 
investigated further. In the next subsections we will look in more detail at the value of the 
radial epicyclic frequency at its maximum.

\subsection{Local extrema of the radial epicyclic frequency}
\label{sec:localextrema}

Of the three frequencies describing orbital motion investigated above, the radial epicyclic 
frequency is the only one that is non-monotonic in $r/M$ for small $j$. We will now 
investigate further the radius at which $\nu_{\mathrm{r}}$ has its maximum and the values 
which it takes there. The location of the maximum can be found by solving the equation

\begin{equation}
        \frac{d\nu_r}{dr} = 0.
\end{equation}

\begin{figure}
  \includegraphics[width=0.98\textwidth]{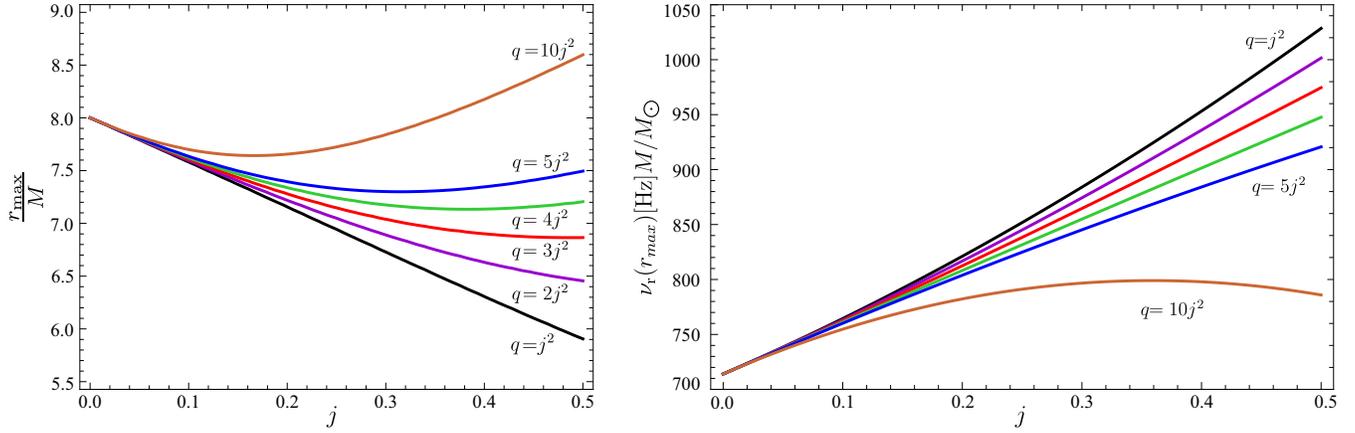}

  \caption{Left: the radius at which the profile of the radial epicyclic frequency 
$\nu_{\mathrm{r}}$ has its maximum. Right: the value of $\nu_{\mathrm{r}}$ at 
$r_\mathrm{max}$.}
  \label{fig:extreme}
\end{figure}

The solution can be found analytically, and is
\begin{eqnarray}
       r_\mathrm{max}=8M \left[ 1 - \frac{47}{64 \sqrt{2}} j + 
\frac{-12596207 + 43683840 \ln\frac{4}{3}}{147456} j^2 - 
\frac{5}{576}\left(-9835 + 34128 \ln \frac{4}{3}\right)q \right]
\end{eqnarray}

The radial epicyclic frequency at $r_\mathrm{max}$ is then given by
\begin{eqnarray}
    \nu_{\mathrm{r}}(r_\mathrm{max})=\sqrt{\frac{c^3}{64\sqrt{2} \pi G M} 
\left[1+ \frac{15}{16 \sqrt{2}} j + \frac{-3972109 + 13824000 \ln\frac{4}{3}}{8192} j^2 
+\frac{5}{16}q\left(1553 - 5400 \ln\frac{4}{3}\right) \right]},
    \label{eq:RadialEpicFreqAtRmax}
\end{eqnarray}
or, in a simplified form,
\begin{eqnarray}
    \nu_{\mathrm{r}}(r_\mathrm{max})=\sqrt{\frac{714.011}{M} (1+ 0.663 j + 0.587 j^2 - 0.15 q)}.
    \label{eq:RadialEpicFreqAtRmaxSimplified}
\end{eqnarray}

The left panel of Figure \ref{fig:extreme} shows the locations of the maximum 
$r_\mathrm{max}$ plotted as a function of $j$, and the right panel shows the values of 
$\nu_{\mathrm{r}}$ at $r_\mathrm{max}$. One can see that $q/j^2$ is having an impact rather 
similar to that in the case of the marginally stable circular orbit, with a minimum appearing 
in the curves for larger values of $q/j^2$ and moving to smaller values of $j$ as $q/j^2$ is 
increased. From the right panel, one sees that the largest values of $q/j^2$ may even cause 
$\nu_{\mathrm{r}}(r_\mathrm{max})$ to become a decreasing function of $j$ at high enough 
rotation speeds.

\subsection{Precession frequencies}

For matter orbiting around neutron stars, the quantities that may be directly observable 
could be related to combinations of the epicyclic frequencies, rather than to the frequencies 
themselves. In particular, the precession frequencies (periastron or nodal) may play an 
important role.

The periastron precession frequency is given by
\begin{eqnarray}
      \nu_{\mathrm{P}}^{\mathrm{r}}=\nu_K-\nu_r,
      \label{eq:periastron}
\end{eqnarray}
while the nodal precession frequency is given by
\begin{eqnarray}
      \nu_{\mathrm{P}}^{\mathrm{\theta}}=\nu_K-\nu_\theta,
      \label{eq:nodal}
\end{eqnarray}
with $\nu_K$, $\nu_r$ and $\nu_\theta$ being given by equations (\ref{eq:Kepler}), 
(\ref{eq:epicrad}) and (\ref{eq:epicvert}).

\begin{figure*}
  \begin{center}
  \includegraphics[width=0.98\textwidth]{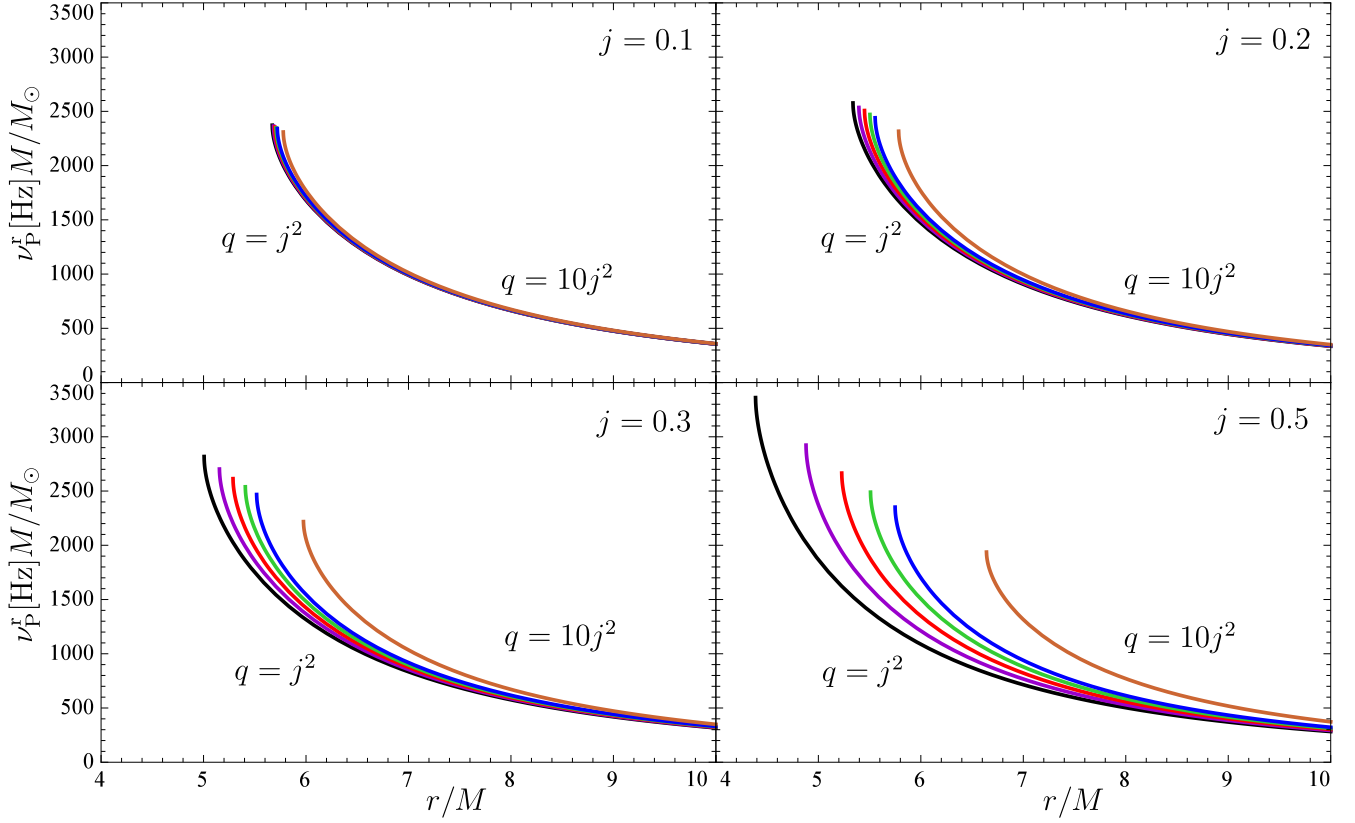}
  \caption{The radial profile of the periastron precession frequency 
$\nu_{\mathrm{P}}^{\mathrm{r}}$.}
  \label{fig:periastron}
  \end{center}
\end{figure*}

\begin{figure*}
  \begin{center}
  \includegraphics[width=0.98\textwidth]{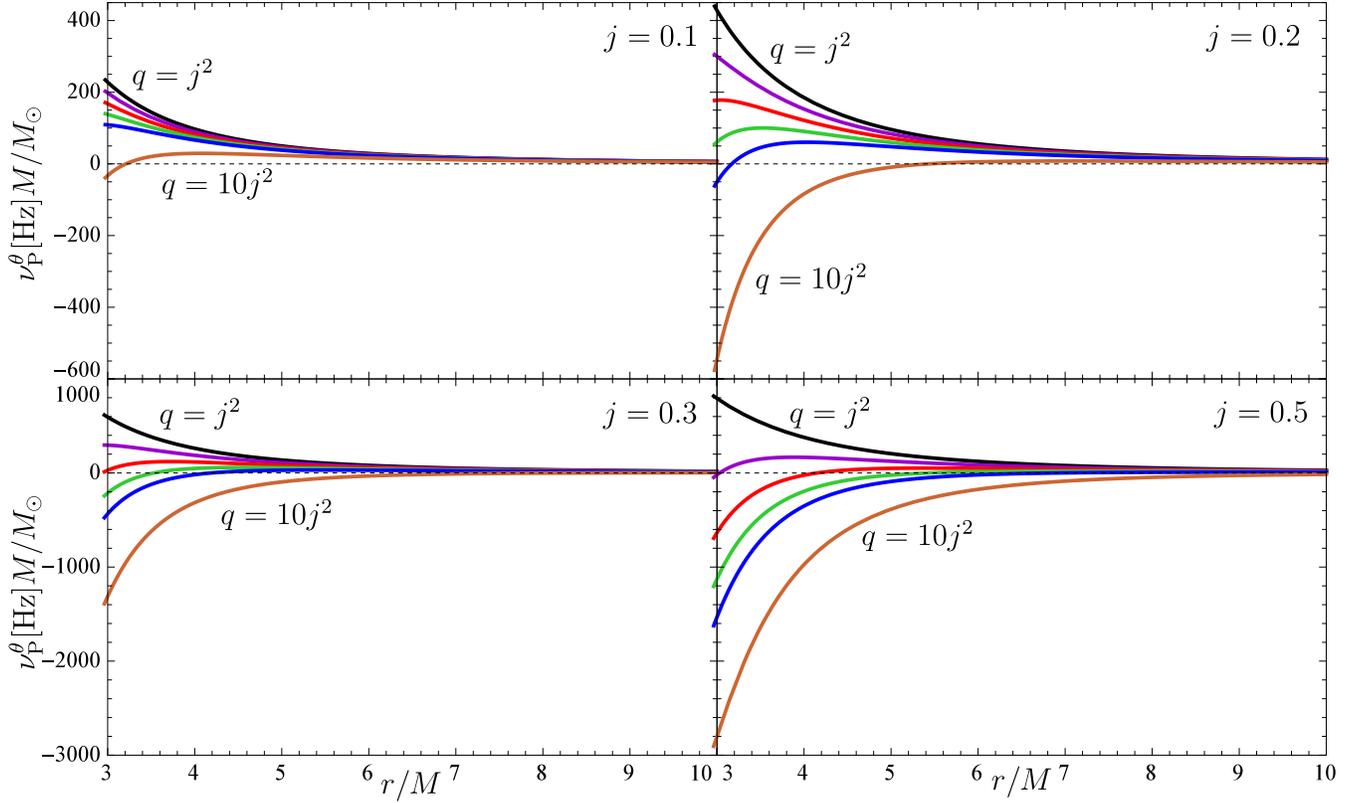}
  \caption{The radial profile of the nodal precession frequency 
$\nu_{\mathrm{P}}^{\mathrm{\theta}}$.}
  \label{fig:nodal}
  \end{center}
\end{figure*}

For calculation of precession frequencies, we again chose $j = 0.1, 0.2, 0.3, 0.5$ and we 
varied the quadrupole parameter $q$ in the range $q/j^2 = 1,2,3,4,5,10$ for each value of 
$j$. Results are plotted in Figures \ref{fig:periastron} and \ref{fig:nodal}. In both cases, 
the impact of $q/j^2$ becomes progressively more significant as $j$ increases. Note that the 
nodal precession frequency can become negative, since the radial epicyclic frequency can be 
higher than the orbital frequency. For low values of $j$, this happens only if $q/j^2$ is 
large and then only very near to $r=3M$, while for larger values of $j$ it happens also for 
smaller $q/j^2$.

\subsection{Comparison of the Keplerian frequency and the vertical epicyclic frequency}

\begin{figure*}
  \begin{center}
  \includegraphics[width=0.98\textwidth]{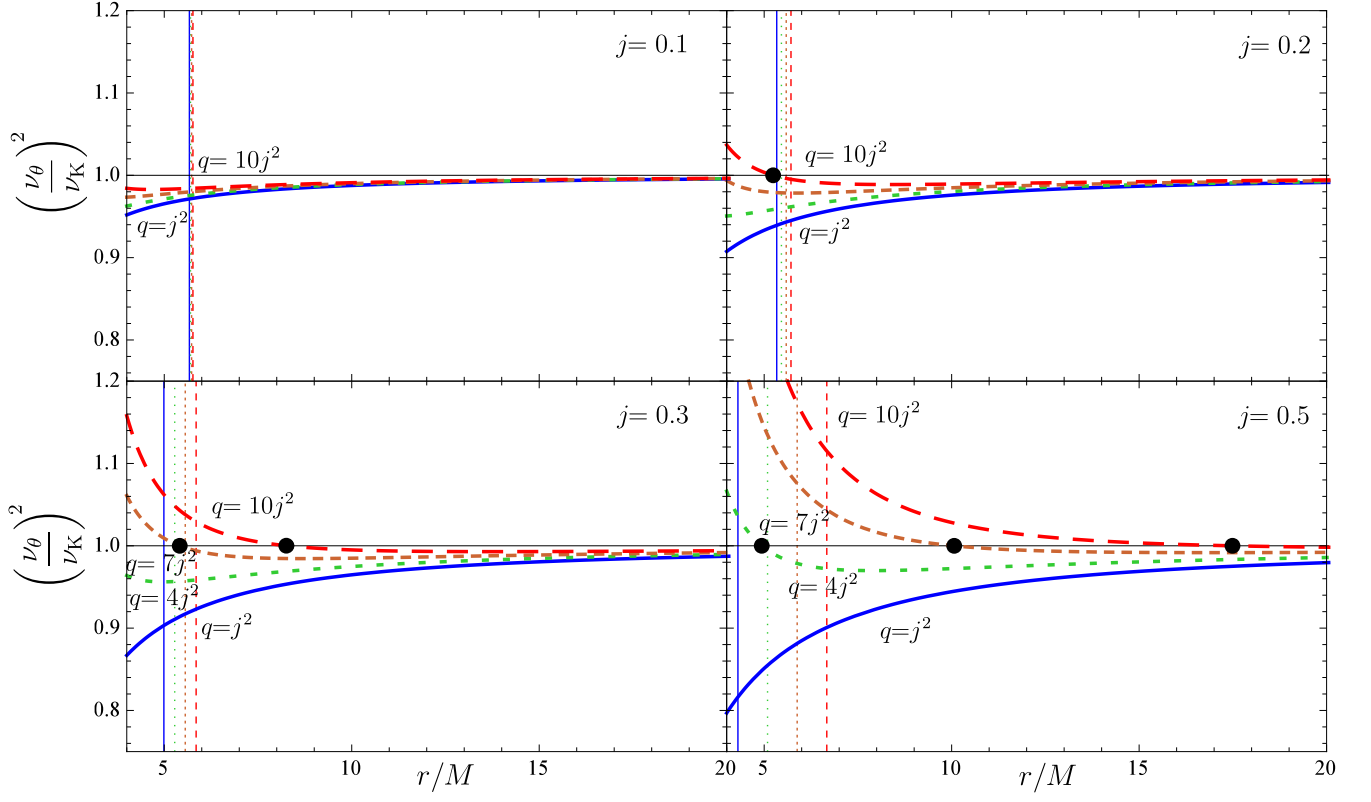}
  \caption{Points where the Keplerian and vertical epicyclic frequencies are equal.}
  \label{fig:ratioThetaKepler}
  \end{center}
\end{figure*}

As mentioned previously, within Newtonian theory it has been found by 
\cite{Gon-Klu-Ste-Wis:2014:PHYSR4:} that the vertical epicyclic frequency can be larger than 
the frequency of orbital motion if the central object is sufficiently oblate and we have seen 
that this can happen also for neutron stars (in general relativity) even if they are {\em 
not} very non-spherical. Here, we calculate the ratio of vertical epicyclic frequency to 
Keplerian frequency of orbital motion, and investigate when it is larger than one. The square 
of the ratio of these two frequencies can be written as
\begin{eqnarray}
      \left[\frac{\nu_{\theta}(r;M,j,q)}{\nu_{K}(r;M,j,q)}\right]^2=1-jK(r)+j^2L(r)-qM(r),
      \label{eq:ratioThetaKepler}
\end{eqnarray}
where
\begin{eqnarray}
      K(r)&=&4M^{3/2}r^{-3/2}, \nonumber\\
      L(r)&=&\frac{3}{8}M^{-2}r^{-2}(8M^4+35M^2r^2-30Mr^3+15r^4)+N(r), \nonumber\\
      M(r)&=&\frac{15}{8}M^{-2}(7M^2-6Mr+3r^3)+N(r), \nonumber\\
      N(r)&=&\frac{45}{16}M^{-3}(M-r)(2M^2-2Mr+r^2)\ln\left(\frac{r}{r-2M}\right). \nonumber
\end{eqnarray}
We plot the results in Figure \ref{fig:ratioThetaKepler}, marking the cases where the 
Keplerian frequency is equal to the vertical epicyclic frequency. These points correspond to 
situations where the relativistic nodal frequency vanishes and changes sign.

The position where the nodal frequency vanishes needs to be compared with that of the 
marginally stable circular geodesic for the values of $j$ and $q/j^2$ being used, because the 
effect of nodal frequency switching is relevant only in the regions where the circular 
geodesic motion is stable against perturbations, i.e., outward of the marginally stable 
circular orbit at $r_\mathrm{ms}$. We have calculated the locations of $r_\mathrm{ms}$ using 
the relation (\ref{eq:Rms}), and these are shown as vertical lines in Figure 
\ref{fig:ratioThetaKepler} with the line-style corresponding to that for the 
$\nu_\theta/\nu_\mathrm K$ profile with same value of $q/j^2$. We can see some cases 
(especially for low values of $j$) where $\nu_\theta/\nu_\mathrm K=1$ occurs at a radius 
below $r_\mathrm{ms}$. On the other hand, for higher values of $j$ and $q/j^2$, $\nu_\theta > 
\nu_\mathrm K$ can occur also for orbits that are stable with respect to radial 
perturbations.

\begin{figure*}
  \begin{center}
  \includegraphics[width=0.50\textwidth]{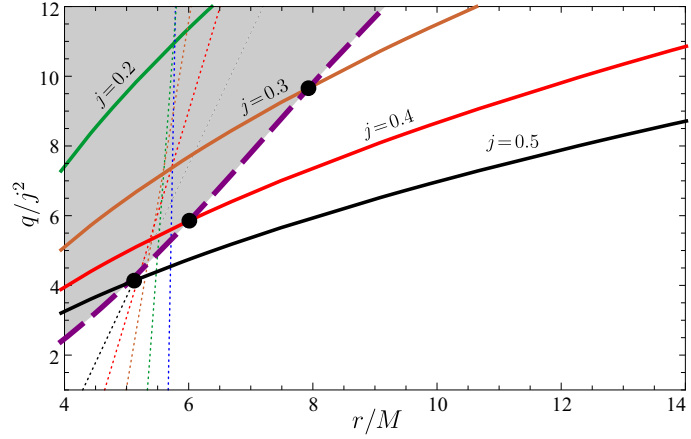}
  \caption{The value of $q/j^2$ giving $\nu_\theta = \nu_\mathrm K$ is plotted as a function 
of $r/M$ for various values of $j$. The dotted lines show the value of 
$r_\mathrm{ms}$ for each $j$ and the purple dashed line marks the stellar surface.}
  \label{fig:QFromRatioThetaKepler}
  \end{center}
\end{figure*}

We now investigate the locations where
\begin{eqnarray}
\left[\frac{\nu_{\theta}(r;M,j,q)}{\nu_{K}(r;M,j,q)}\right]^2=1.
      \label{eq:ratioThetaKepler2}
\end{eqnarray}
Since $\nu_\theta$ is always equal to $\nu_\mathrm K$ in the non-rotating Schwarzschild case, 
one cannot find the location for the nodal switching point as an expansion around the 
Schwarzschild value. The values of $q/j^2$ satisfying (\ref{eq:ratioThetaKepler2}) are given 
by
\begin{eqnarray}
\frac{q}{j^2} = 1 + \frac{64 M^{9/2} \sqrt{r}-48 j M^5}{15 j r^2 
\left[-2 M \left(7 M^2-6 M r+3 r^2\right)-
3 (M-r) \left(2 M^2-2 M r+r^2\right) \ln \left(\frac{r}{r-2 M}\right)\right]},
\end{eqnarray}
and in Figure \ref{fig:QFromRatioThetaKepler} we plot these against $r/M$ for $j = 
0.2, 0.3, 0.4, 0.5$. The dotted lines show the value of $r_\mathrm{ms}$ for each $j$ and the 
purple dashed line indicates the stellar surface; we have marked the point on each of the 
$j$ curves beyond which $r$ is greater than both of those limits.


\section{Models of twin HF QPOs and the frequency ratios in the Hartle--Thorne geometry}
\label{sec:models}

In the previous section we have investigated the behavior of the orbital, epicyclic and 
precession frequencies. We have seen that as $j$ increases, the role of the quadrupole term 
becomes increasingly important. In this section, we focus on particular combinations of 
frequencies that play key roles for models of QPOs - the twin peaks observed in power spectra 
of LMXBs. The origin of these peaks is still under investigation, but several candidate 
models are using frequencies associated with orbital motion around the central objects.

\subsection{Frequency ratios of oscillatory modes for the twin HF QPO models}

In this section we study the behavior of the frequencies of oscillatory modes given by some 
current models for the twin HF QPOs observed in atoll and Z-sources, where a neutron star is 
accreting matter from a low-mass companion (for the differences between atoll and Z-sources 
see \citet{Has-Kli:1989:AA:}). We consider the specific models presented in 
Table~\ref{tab:modely}, where we also give the combinations of frequencies that are 
attributed to the upper and lower observed peaks in each case. Our selection of models is 
based on \cite{2011:Torok:Kotr:Sram:Stu:}. None of the models is currently uniquely 
preferred, because each of them has (different) theoretical problems. The RP (Relativistic 
Precession) model \citep{Ste-Vie:1999:PHYSRL:} considers relativistic epicyclic motion of 
blobs at various radii in the inner parts of the accretion disc; The TD (Tidal Disruption) 
model \citep{2008:Cad-Cal-Kost:} proposes that the QPOs are generated by a tidal disruption 
of large accreting inhomogeneities; The WD (Warped Disk) model \citep{Kat:2001} considers a 
somewhat exotic disc geometry that causes a doubling of the observed lower QPO frequency; The 
ER (Epicyclic Resonance) model \citep{abr-klu:2001} involves different combinations of 
axisymmetric disk-oscillation modes; RP1~\citep{2005:Bursa:} and 
RP2~\citep{2007:Tor-Stu-Bak:} use different combinations of non-axisymmetric disk-oscillation 
modes. We are considering here only geodesic oscillation models where the oscillatory modes 
are combinations of the orbital and epicyclic frequencies of near-circular geodesic motion in 
the equatorial plane.

\begin{table*}
  \caption{An overview of the orbital models of QPOs, with specification of the 
upper~$\nu_{\mathrm{U}}$ and lower~$\nu_{\mathrm{L}}$ frequencies.}
  \label{tab:modely}
  \begin{tabular*}{\textwidth}{@{\extracolsep{\fill}}lllllll@{}}
  \hline\hline
  \noalign{\smallskip}
   Frequency&RP&TD&WD&RP1&RP2&ER\\
  \noalign{\smallskip}\hline\noalign{\smallskip}
  $\nu_{\mathrm{U}}$ & $\nu_{\mathrm{K}}$ &
  $\nu_{\mathrm{K}}+\nu_{\mathrm{rad}}$ &
  $2\nu_{\mathrm{K}}-\nu_{\mathrm{rad}}$ & $\nu_{\theta}$ &
  $2\nu_{\mathrm{K}}-\nu_{\theta}$ & $\nu_{\theta}$ \\

  $\nu_{\mathrm{L}}$ & $\nu_{\mathrm{K}}-\nu_{\mathrm{rad}}$ &
  $\nu_{\mathrm{K}}$ & $2(\nu_{\mathrm{K}}-\nu_{\mathrm{rad}})$ &
  $\nu_{\mathrm{K}}-\nu_{\mathrm{rad}}$ &
  $\nu_{\mathrm{K}}-\nu_{\mathrm{rad}}$ & $\nu_{\mathrm{rad}}$ \\
  \noalign{\smallskip}\hline
  \end{tabular*}
\end{table*}

\begin{figure*}
  \includegraphics[width=0.98\textwidth]{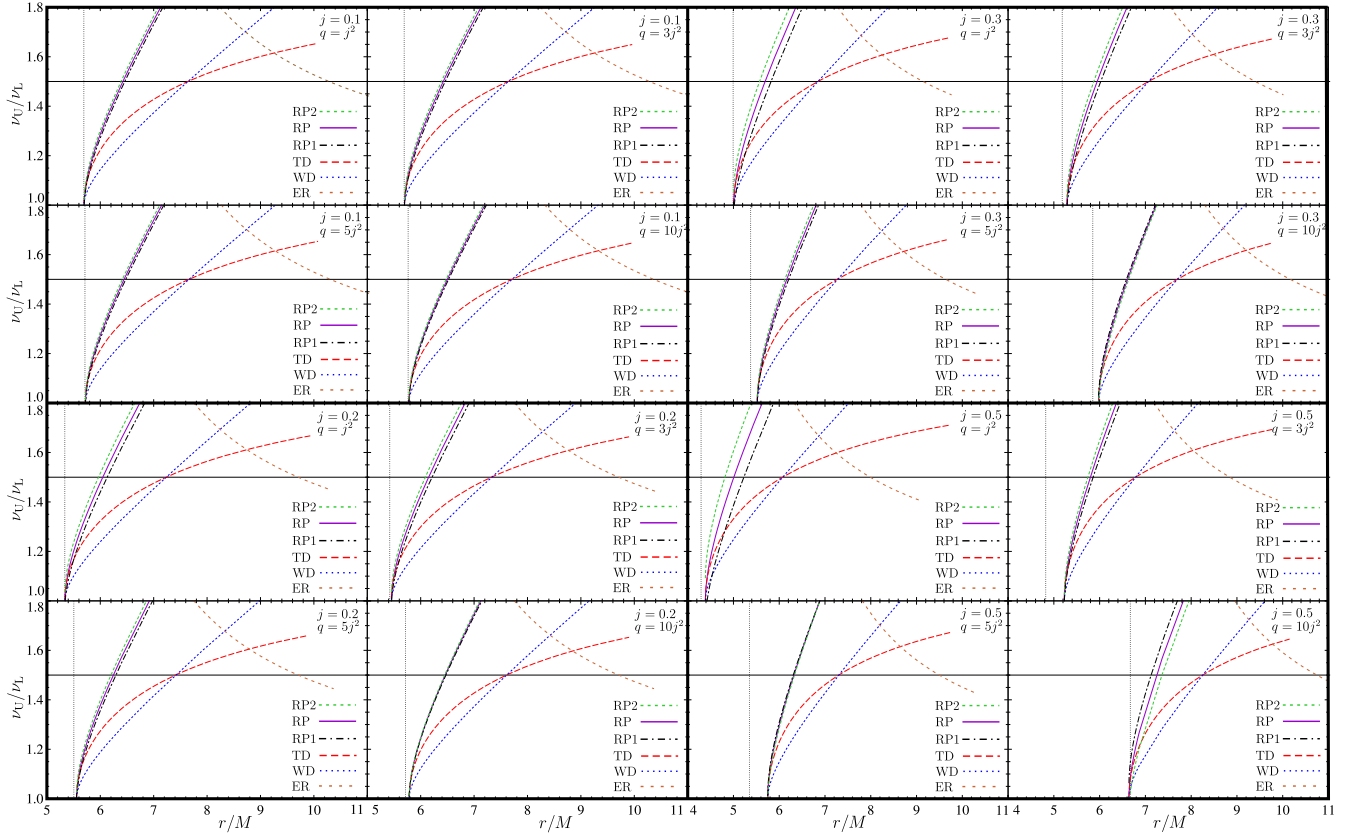}
  \caption{The orbital models of QPOs (with $j=0.1,~0.2,~0,3,~0.5$). The dependence of 
$\nu_{\mathrm{U}}/\nu_{\mathrm{L}}$ on $r/M$ is shown. The vertical dotted line indicates the 
location of the marginally stable orbit $r_{\mathrm{ms}}$, as given by equation 
(\ref{eq:Rms}), and the horizontal line indicates the ratio 
$\nu_{\mathrm{U}}~:~\nu_{\mathrm{L}}~=~3~:~2$.}
  \label{fig:models}
\end{figure*}

In Figure \ref{fig:models} we show the radial profiles of the frequency ratio $\nu_\mathrm 
U/\nu_\mathrm L$ for all of the six QPO models being considered here.

As in the previous section, we use the values $j = 0.1, 0.2, 0.3, 0.5$, which includes the 
maximal range of the spin parameter $j$ for the Hartle--Thorne approximation 
\citep{Urb-Mil-Stu:2013:MONNR:}, and for the quadrupole parameter $q$ we use values obtained 
from taking $q/j^2 = 1,3,5,10$, covering the most astrophysically relevant models for 
rotating neutron stars. It can be seen that for the RP models, the ratio $\nu_\mathrm 
U/\nu_\mathrm L=3/2$ occurs just slightly outside the location of the marginally stable 
orbit, meaning that the QPOs will generally be related to the region very close to the inner 
edge of the accretion disk, but for the TD and WD models the radius is larger and it is even 
more so for the ER model. However all of the considered models are giving the radius where 
$\nu_\mathrm U/\nu_\mathrm L=3/2$ as being between $r_\mathrm{ms}$ and $r=11M$.

We also note that the RP, RP1 and RP2 models give extremely similar profiles, with only very 
minor differences. The profiles of the WD and TD models are different, but they always cross 
each other at the frequency ratio $3:2$. For the ER model, the ratio diverges at the 
marginally stable orbit, since the radial epicyclic frequency has to vanish there, and this 
is the only case where the ratio decreases with increasing radius. In all of the models, the 
radial profiles depend only slightly on the quadrupole parameter for small values of $j<0.2$, 
but the dependence increases with increasing $j$, and for $0.3<j<0.5$ it can be very strong, 
especially close to the marginally stable orbit.

\begin{figure}
  \includegraphics[width=0.98\textwidth]{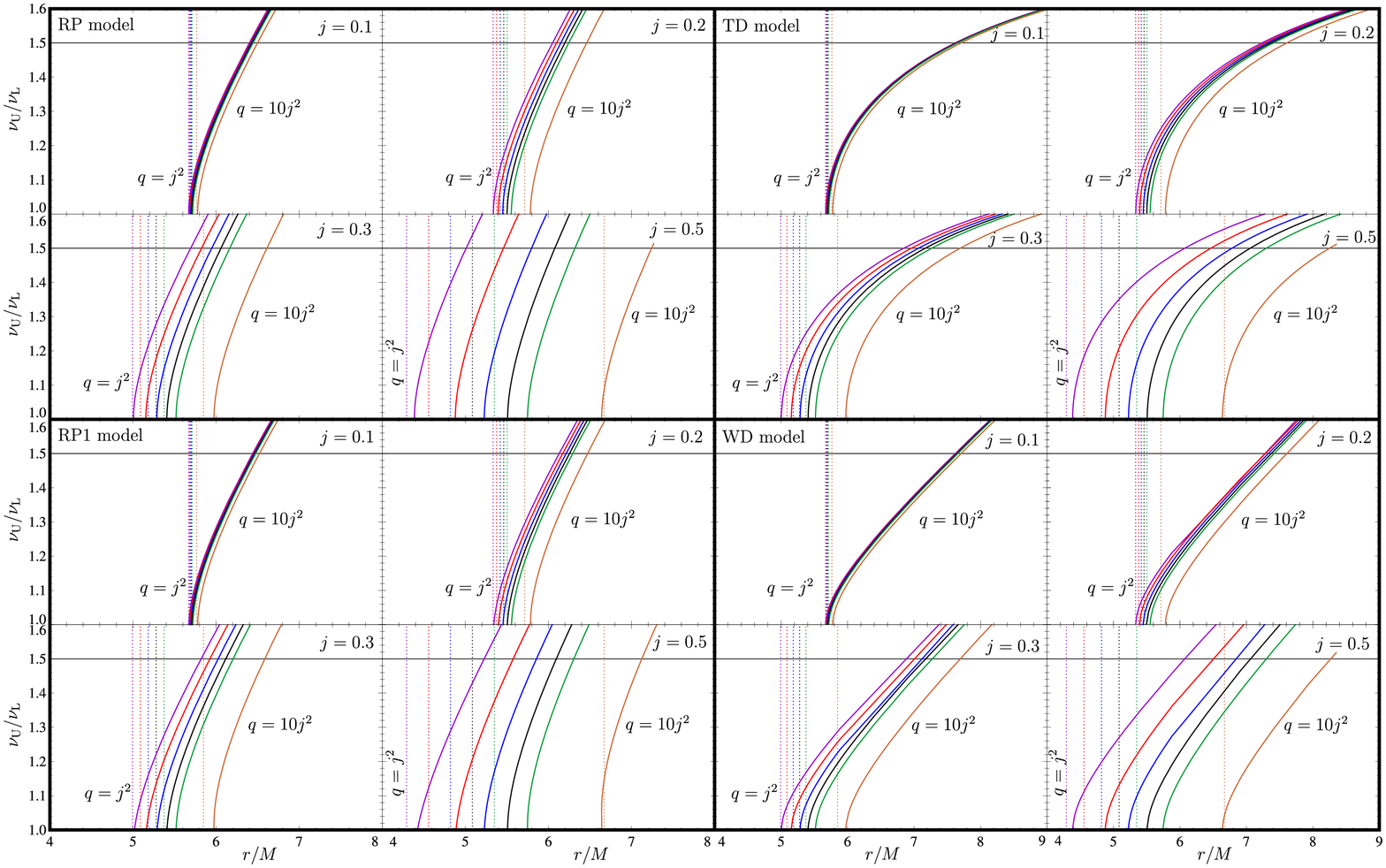}
  \includegraphics[width=0.98\textwidth]{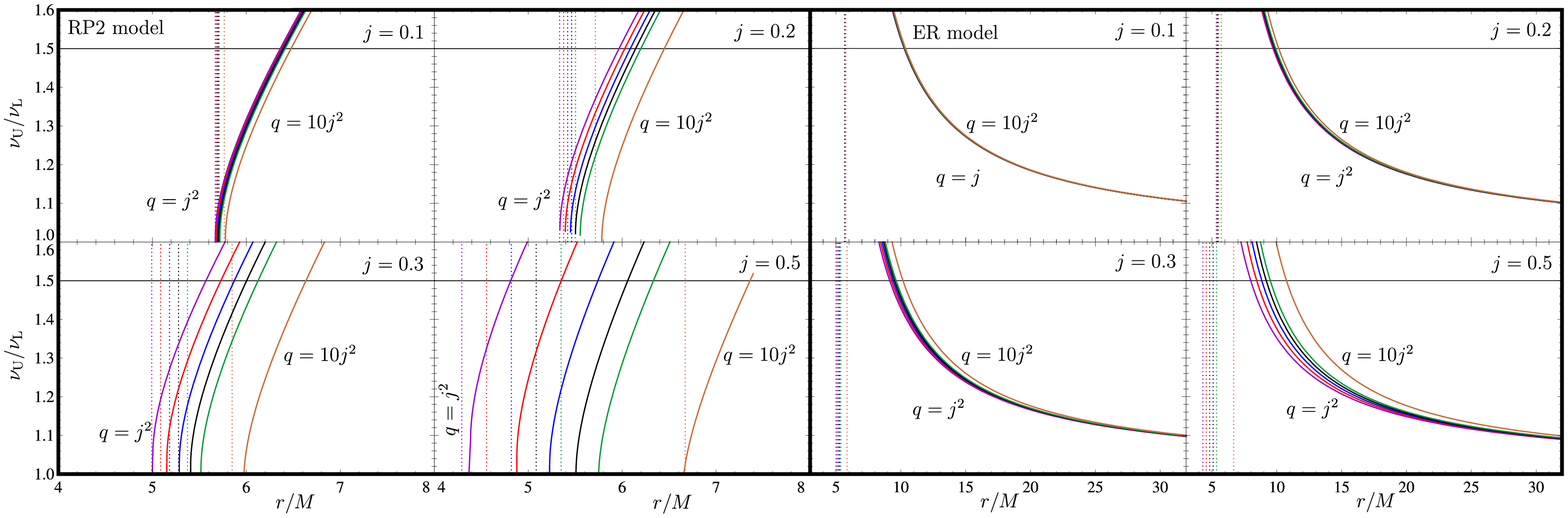}
  \caption{Individual results for each of the orbital QPO models: RP, RP1, RP2, TD, WD, ER 
(top left to bottom right). The ratio $\nu_{\mathrm{U}}/\nu_{\mathrm{L}}$ is plotted against 
$r/M$ for each model. The vertical dotted lines indicate the location of the marginally 
stable orbit $r_{\mathrm{ms}}$ for each value of $q$.}
  \label{fig:modelRP}
\end{figure}

In Figure \ref{fig:modelRP}, we show these results plotted for each of the QPO models 
separately, using $j = 0.1, 0.2, 0.3, 0.5$, as before, and $q/j^2 = 1, 2, 3, 4, 5, 10$. Here 
one can appreciate better the role of $j$ and $q/j^2$ for varying the position of marginally 
stable orbit. Note the qualitative difference between the results for the ER model and the 
others.

\subsection{The frequency ratio $3:2$ in the twin HF QPO models}

\begin{figure}
  \includegraphics[width=0.49\textwidth]{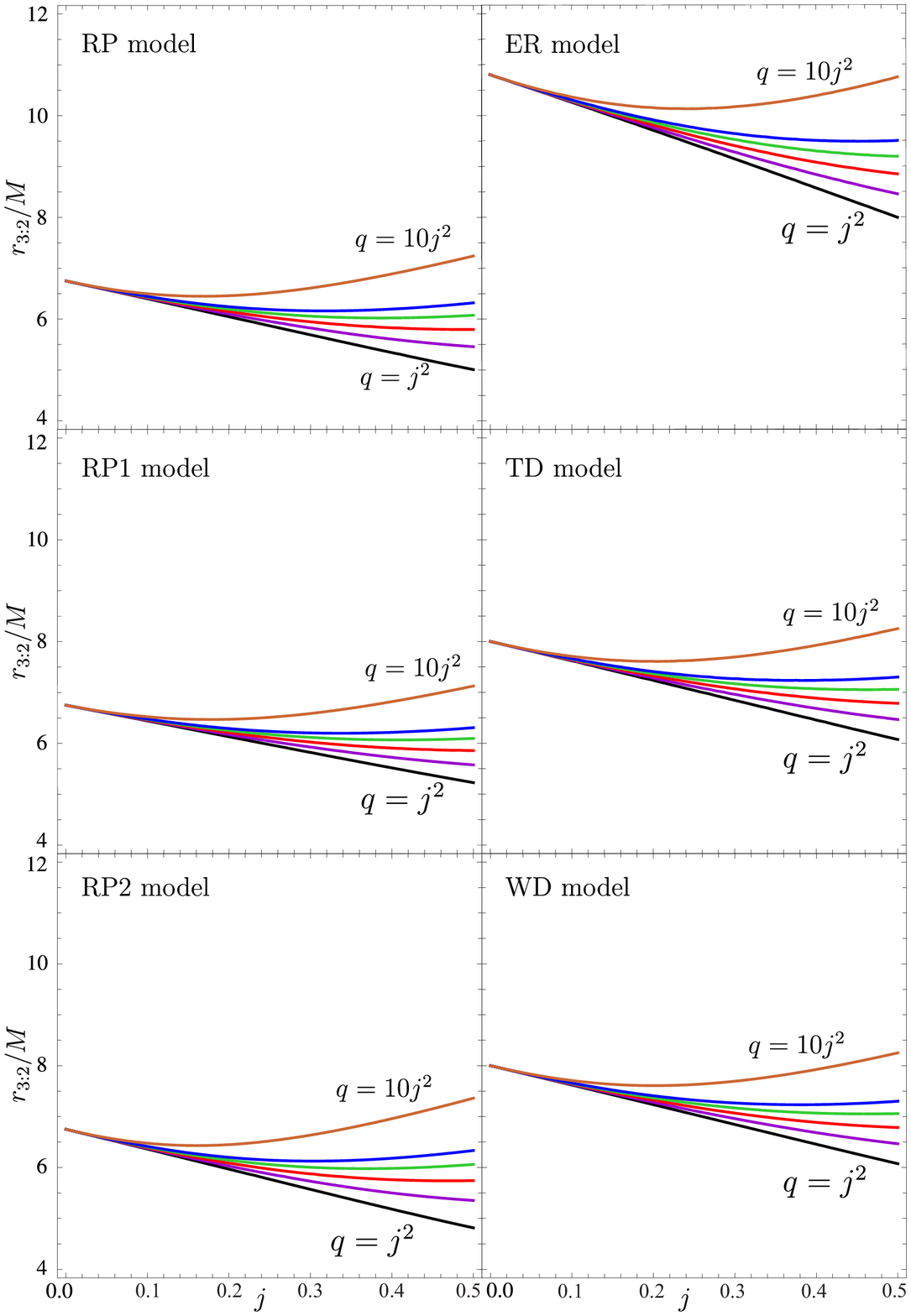}
  \includegraphics[width=0.49\textwidth]{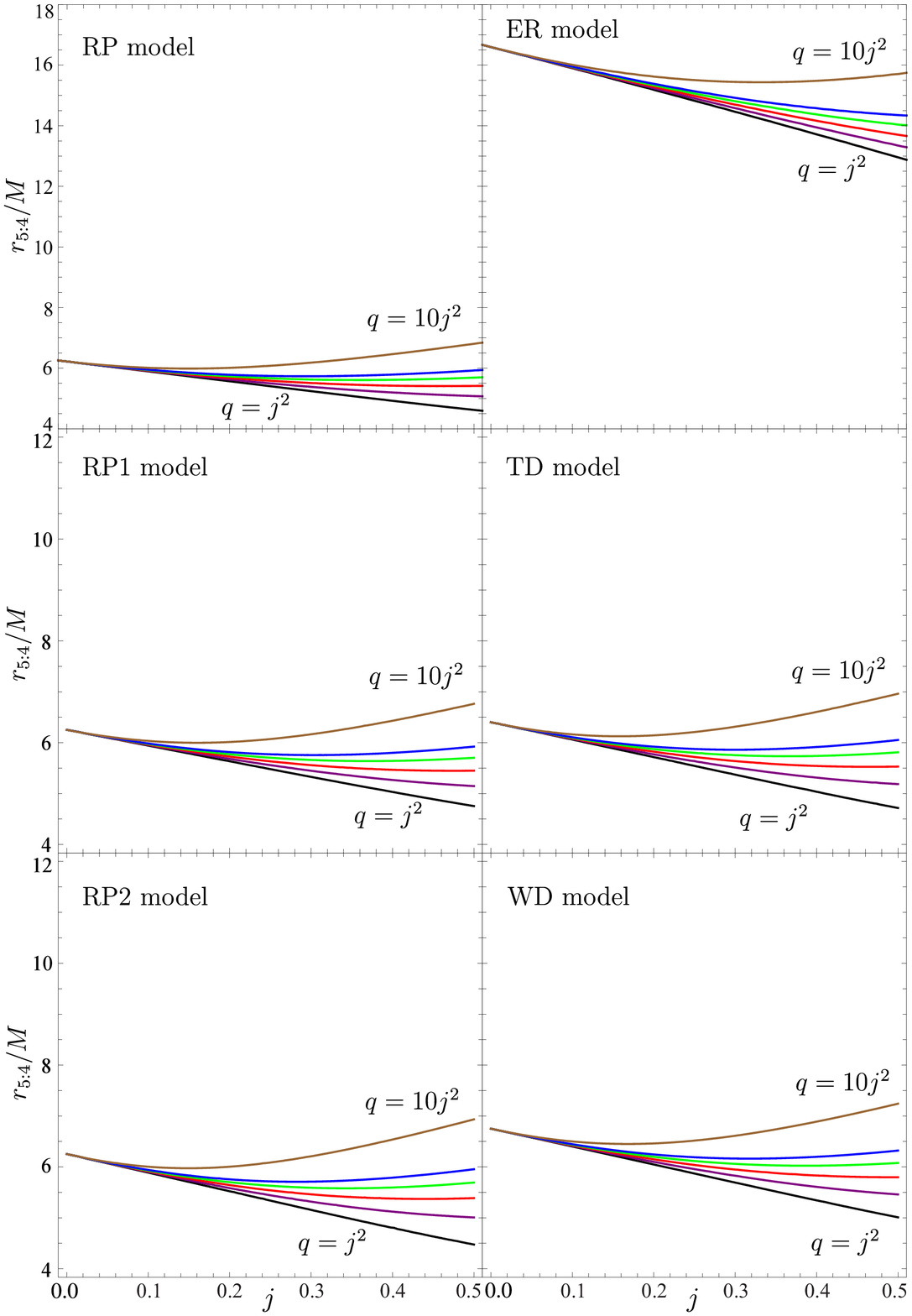}
  \caption{The locations where the resonant QPO ratios 3:2 and 5:4 are generated: For each of 
the models, $r_{3:2}/M$ (left panels) and $r_{5:4}/M$ (right panels) are plotted against $j$ 
for $q/j^2 = 1, 2, 3, 4, 5, 10$.}
  \label{fig:graf:models3over2}
\end{figure}

Observational data for QPOs demonstrates the importance of the frequency ratio $3:2$ 
\citep{Bou-etal:2010:MONNR} and so we have calculated how the location where this ratio is 
generated varies as a function of $j$. The left panels of Figure \ref{fig:graf:models3over2}, 
show this for each of our QPO models for $q/j^2 = 1,2,3,4,5,10$. Note that the behavior of 
$r_{3:2}/M$ demonstrates a qualitatively similar behavior to that for the location of the 
marginally stable orbit: for small values of $q/j^2$, the resonance radius decreases with 
increasing spin, as for Kerr black holes, but for large enough $q/j^2$ it reaches a minimum 
and then increases again.

The right panels of Figure \ref{fig:graf:models3over2} show equivalent results for the ratio 
5:4, which are quite similar.

\section{Comparison of results from the Hartle--Thorne approach and the Lorene/nrotstar 
numerical code}

In this section we test the range of validity of the Hartle--Thorne approach by comparing 
results obtained with it against ones given by the publicly available Lorene/nrotstar code 
\citep{Lorene}. We compare models having the same gravitational mass for a static 
non-rotating star $M_0$ and then keep constant the value of the central pressure (and the 
energy or enthalpy density) when we move to rotating models. We use the APR equation of state 
\citep{APR} for making the comparison.

We focus on the quantity $\Delta f_\mathrm{ISCO}= f_\mathrm{ISCO}-f_\mathrm{ISCO} 
(f=0Hz)$, where $f_\mathrm{ISCO}$ is here the frequency at the innermost stable orbit (either 
the marginally stable circular orbit at $r = r_\mathrm{ms}$ or the equatorial 
surface-skimming orbit at $r = R_\mathrm{eq}$ when $R_\mathrm{eq} > r_\mathrm{ms}$, as 
discussed previously (see also \citet{Pap:2015:MNRAS:} for discussion of when the 
surface of the neutron star lies above/below $r_\mathrm{ms}$). 

This quantity is zero in the non-rotating case and its leading order in the 
slow-rotation expansion goes as $f_\mathrm{rot}$ while the order of its first neglected term 
goes as $\mathcal{O} (f^3_\mathrm{rot})$, i.e. the relative error is of order $\mathcal{O} 
(f^2_\mathrm{rot})$. This is also the case for $\nu^{\theta}_\mathrm P$, but the quantities 
$f_\mathrm{ISCO}$ and $\nu_P^r$ have non-zero values for the non-rotating case, i.e. the 
leading term is of $\mathcal{O} (f^0_\mathrm{rot})$, and since the first neglected term is 
again $\mathcal{O} (f^3_\mathrm{rot})$, this means that the relative error is of order 
$\mathcal{O} (f^3_\mathrm{rot})$, one order higher than that for $\nu^{\theta}_\mathrm P$. 
Because of this $\nu^{\theta}_\mathrm P$ is likely to be the quantity most liable to error 
and so it is sensible to use as our key test quantity $\Delta f_\mathrm{ISCO}= 
f_\mathrm{ISCO}-f_\mathrm{ISCO} (f=0Hz)$ which has an error of the same order as the most 
sensitive quantity of interest. This is not a foolproof argument because of different 
coefficients in the expansions of the quantities concerned, but it should be a reasonable 
guide.

\begin{figure}
  \includegraphics[width=0.49\textwidth]{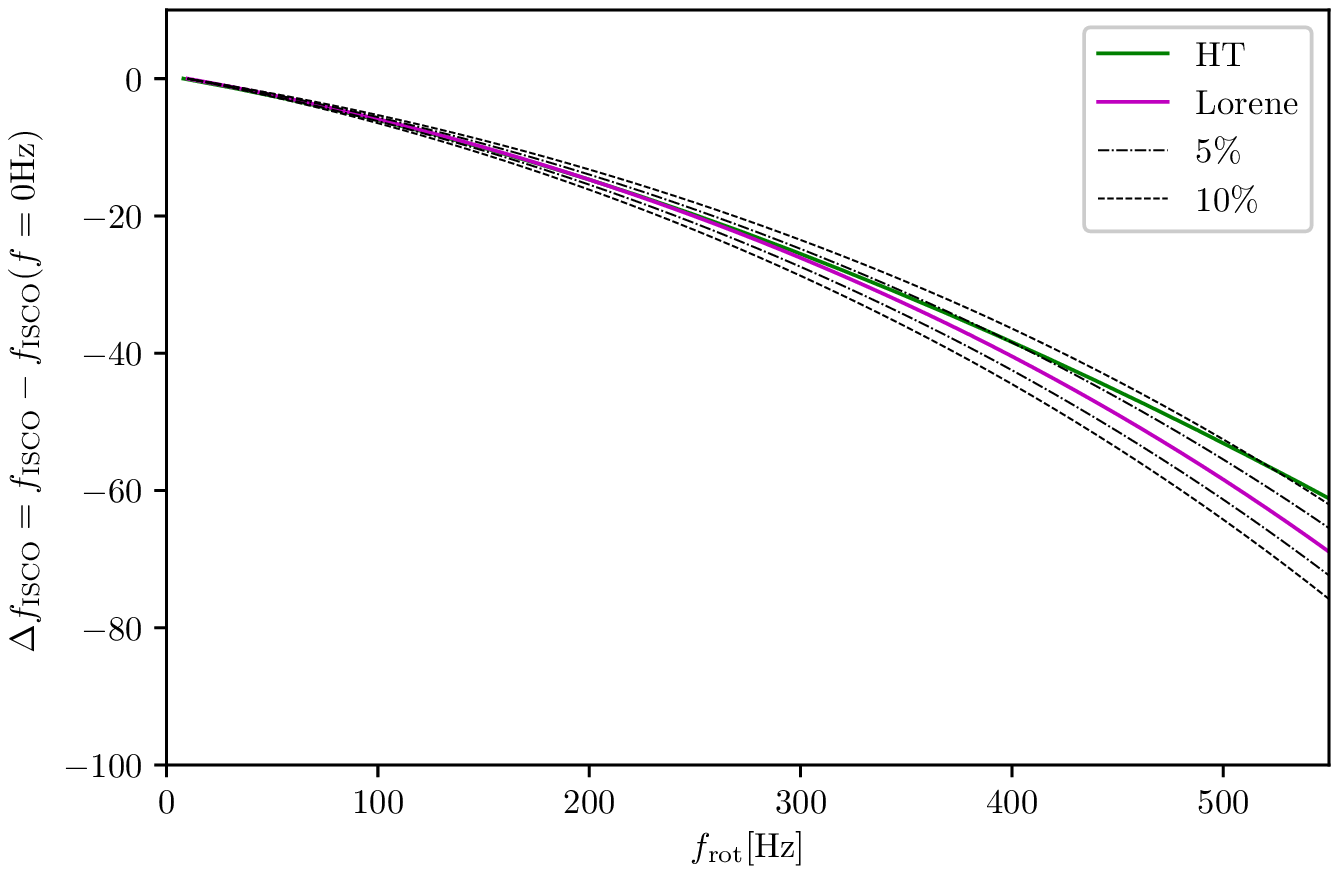}
   \includegraphics[width=0.48\textwidth]{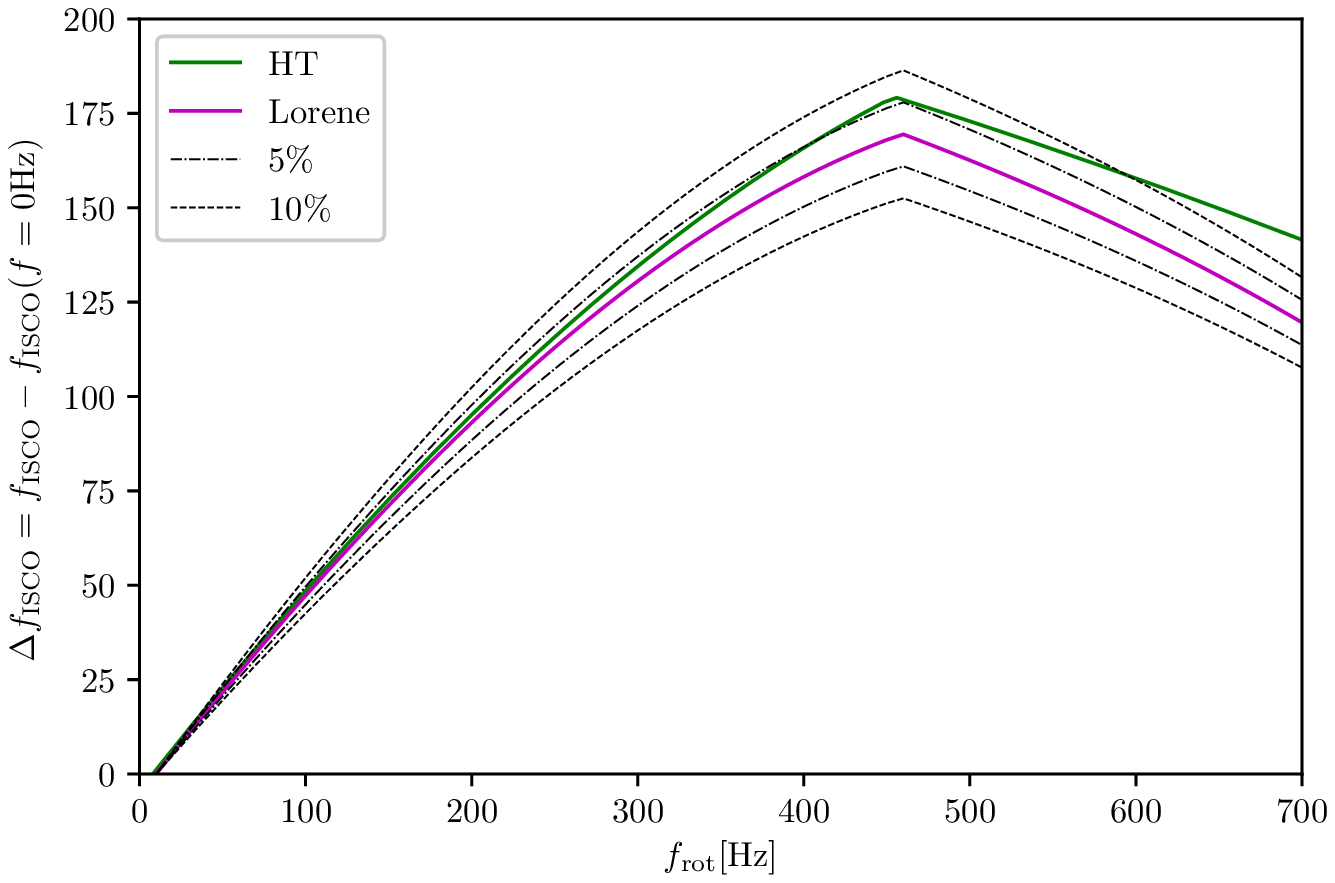}

  \caption{$\Delta f_\mathrm{ISCO}$ is plotted as a function of $f_\mathrm{rot}$ as 
calculated using the Hartle--Thorne and Lorene approaches. The gravitational mass of the 
non-rotating star is $M=1.2M_\odot$ (left) and $M=1.4M_\odot$ (right). See text for details.}
  \label{fig:fiscoM124}
\end{figure}

\begin{figure}
  \includegraphics[width=0.485\textwidth]{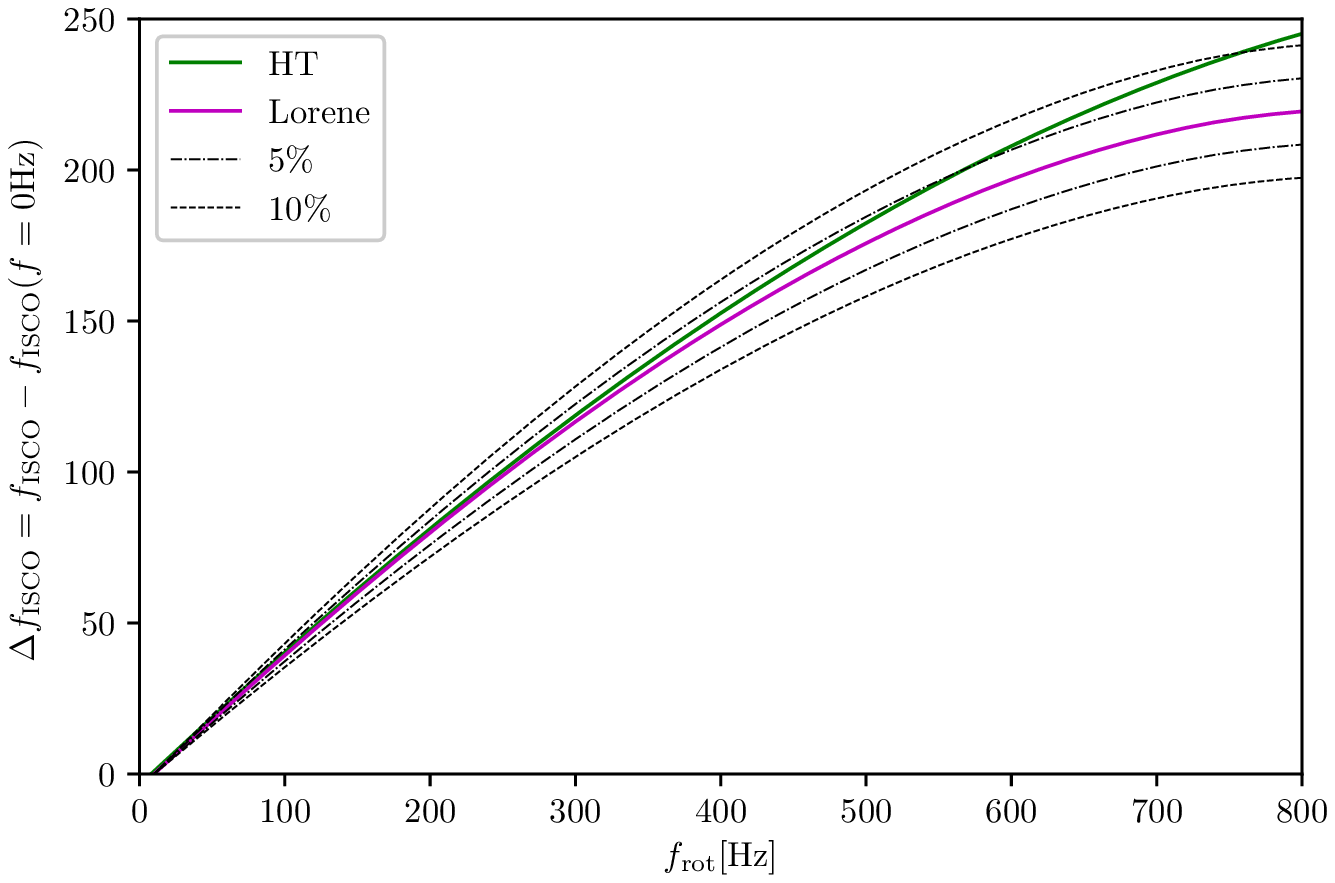}
  \includegraphics[width=0.48\textwidth]{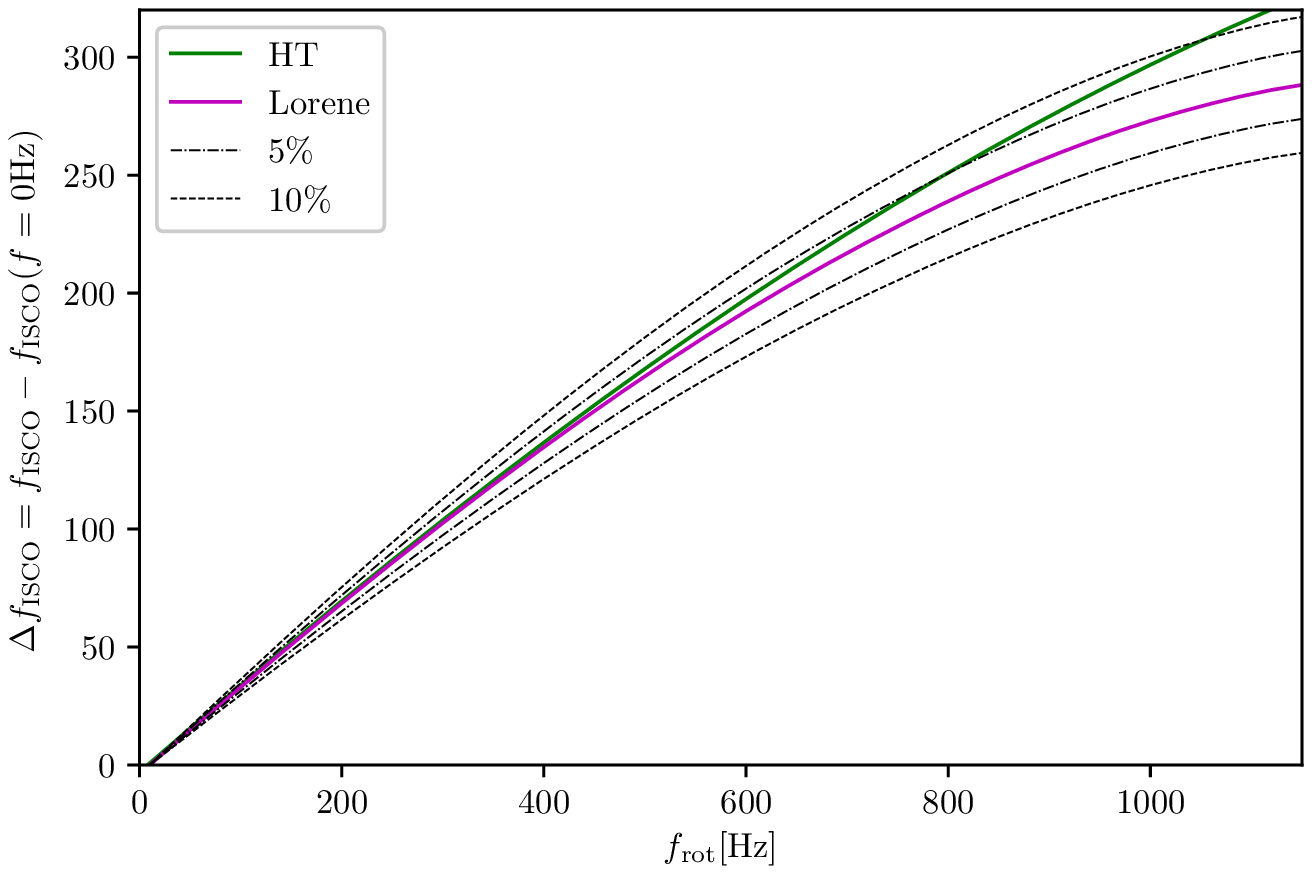}

  \caption{$\Delta f_\mathrm{ISCO}$ is plotted as a function of $f_\mathrm{rot}$ as 
calculated using the Hartle--Thorne and Lorene approaches. The gravitational mass of the 
non-rotating star is $M=1.6M_\odot$ (left) and $M=1.8M_\odot$ (right). See text for details.}
  \label{fig:fiscoM168}
\end{figure}

For the ISCO frequency in the Hartle--Thorne approach, we calculate a gravitational 
mass $M$ for the rotating star, a dimensionless angular momentum $J/M_0^2$ and a 
dimensionless quadrupole moment $Q/M_0^3$, use these to calculate the radius $r_\mathrm{ms}$ 
from eq.\,(\ref{eq:Rms}) and then find the Keplerian frequency at this radius using 
eq.\,(\ref{eq:Kepler}). If $r_\mathrm{ms}$ is less than $R_\mathrm{eq}$, then the latter is 
used instead in eq.\,(\ref{eq:Kepler}). For the gravitational mass of the non-rotating star 
we take $M_0 =1.2,~1.4,~1.6,~1.8 M_\odot$. The rotational frequencies are taken up to the 
frequency where the difference between the Hartle--Thorne and Lorene values for $\Delta 
f_\mathrm{ISCO}$ reaches $\sim 10 \%$. The results of the comparison are shown in 
Figures\,\ref{fig:fiscoM124} and \ref{fig:fiscoM168}, where we show also the 5\% and 10\% 
intervals around the Lorene values. The positive gradients correspond to $r_\mathrm{ms}$ 
being used, and the negative gradients to $R_\mathrm{eq}$. Since we are here dealing with 
specific models, the location of the stellar surface is directly known.

The two approaches are in excellent agreement for the lower rotation frequencies, as 
expected; when differences begin to emerge, Hartle--Thorne is giving higher values for 
$\Delta f_\mathrm{ISCO}$ in all cases. We see that for lower masses the difference between 
the Lorene and Hartle--Thorne values of $\Delta f_\mathrm{ISCO}$ are smaller than 10\% up to 
frequencies of about 500Hz, while for higher masses the rotational frequencies can go up to 
700Hz (for $M=1.6 M_\odot$) or 1000Hz (for $M=1.8 M_\odot$). Since it is generally believed 
that neutron stars in QPO sources have accreted matter from the companion star they should 
have rather high masses.


\section{Summary}

We have examined the behavior of the orbital frequency, the radial epicyclic frequency and 
the vertical epicyclic frequency in the external Hartle--Thorne geometry for rotating compact 
stars. The primary parameters of the geometry are the gravitational mass $M$, the angular 
momentum $J$ and the quadrupole moment $Q$ and we have calculated the corresponding values 
for these from models of neutron stars constructed using realistic equations of state for 
nuclear matter. Our approach can be applied for any future equation of state if the universal 
relations are used as described in Section\,2\footnote{During the refereeing process and 
preparation of the revised version of this manuscript, a paper by \citet{2018LukLin} has been 
published where it is demonstrated that the frequency at the marginally stable orbit around 
rotating neutron stars is also a universal function of the neutron star properties.}. Apart 
from the individual frequencies, we have also investigated the behavior of combinations of 
them which are involved in some models for QPOs observed in X-rays from LMXBs.

We have seen that the dimensionless angular momentum $j$ and dimensionless quadrupole moment 
$q$ play an important role for the behavior of the orbital and epicyclic frequencies, 
especially for the ratio between the vertical epicyclic frequency and the Keplerian 
frequency, becoming particularly important for stars rotating with the higher angular 
velocities. Our investigation was motivated by the observations of QPOs in the X-ray spectra 
of LMXBs and we have investigated the behavior of the frequencies relevant for several of the 
models proposed for explaining these.

It may seem redundant to spend time with more models, but at present we do not know which 
model may be relevant for the sources of QPOs and are not even sure whether only one model 
will be applicable for all of them. If the radiation does come from the accretion disc and is 
modulated by its oscillations, different oscillation modes may well be involved for different 
sources, depending on the geometry of the binary system. For example, if the equatorial plane 
of the compact object is the same as the orbital plane of the binary system, one may expect 
radial perturbations of the disc caused by matter falling in from the donor star. On the 
other hand, if the equatorial plane of the compact star is almost perpendicular to the 
orbital plane of the binary, one may expect vertical perturbations of the disk instead. This 
is something which may well clarify in the near future.

\begin{figure}
  \includegraphics[width=0.95\textwidth]{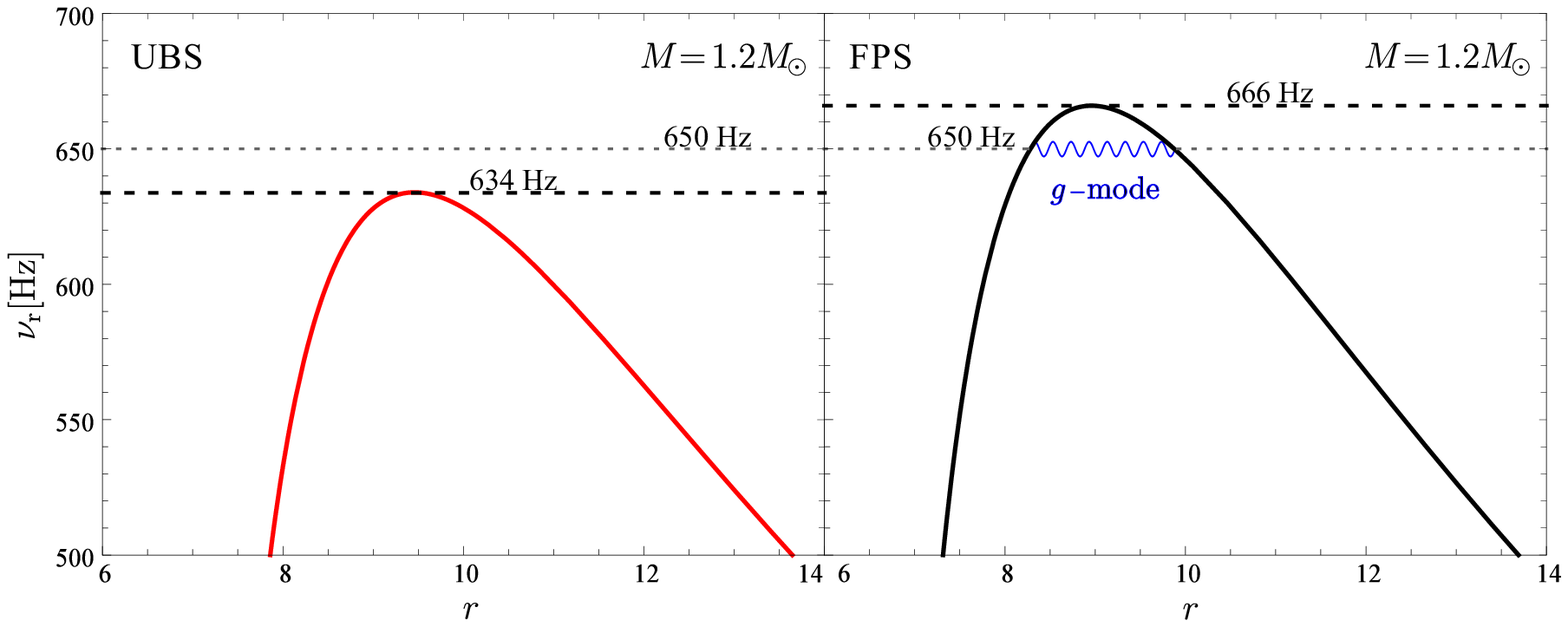}
  \includegraphics[width=0.945\textwidth]{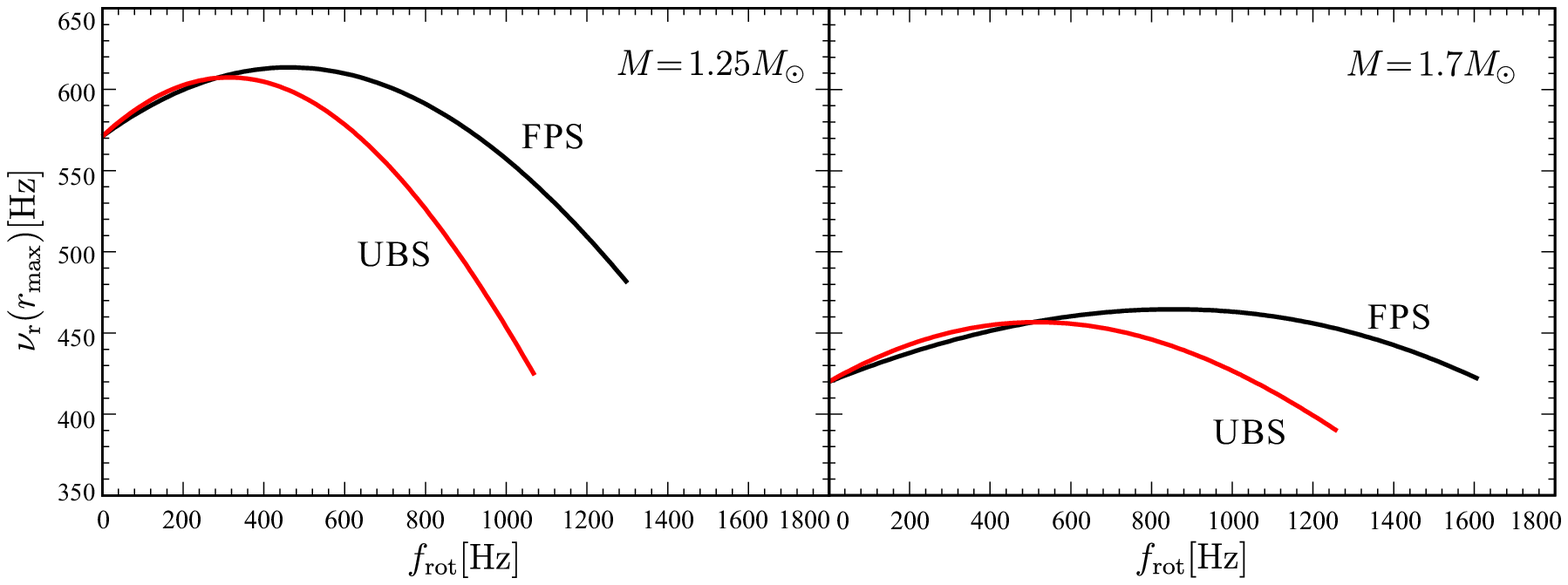}
  \caption{Upper: the radial profile of the radial epicyclic frequency $\nu_{\mathrm{r}}$ for 
stars with gravitational mass $1.2~M_\odot$ rotating with frequency of 580~Hz for two 
selected equations of state. The dashed horizontal lines indicate the values of 
$\nu_{\mathrm{r}}(r_\mathrm{max})$ determining the trapping of discoseismic g-modes. The 
dotted horizontal line indicates a particular value of the trapped g-mode frequency at 666~Hz 
which is allowed for the FPS EoS but not for UBS. Lower: the maximum radial epicyclic 
frequency, calculated using equation (\ref{eq:RadialEpicFreqAtRmax}), plotted as a function 
of the stellar rotational frequency for stars with gravitational mass $1.25~M_\odot$ (left) 
and $1.7~M_\odot$~(right).}
  \label{fig:gmode}
\end{figure}

Finally we note that the results presented here can be relevant not only for the particular 
models mentioned in Section \ref{sec:models} but also for a wider class of orbital models. 
For instance, the maximum of the radial epicyclic frequency determines propagation of some 
discoseismic modes \citep{kat-etal:1998:Kyoto:,wag-etal:2001} and so there would be an 
application for making estimates of stellar spin rates based on association of QPOs with 
oscillations trapped near to the inner edge of thin accretion discs 
\citep{Kato:1989:PASJ:,Kat:2001}. This is illustrated in Figure \ref{fig:gmode} where we show 
the dependence of the maximum value of the radial epicyclic frequency 
$\nu_{\mathrm{r}}(r_\mathrm{max})$ for compact and highly oblate stars. Similar examples can 
be found for the vertical epicyclic frequency and models dealing with the Lense-Thirring 
precession effect \citep{Ingram-Done-Fragile:2009:MNRAS:,Ros-Klu:2014,Wis-etal:RAG:2015:,
ingram-etal:2015:ApJ:,:Tsang-Pappas:2016:ApJ:}. The formulas given in Section 3 for 
determining the epicyclic frequencies (and the ratio between the vertical epicyclic frequency 
and the Keplerian frequency) can therefore be used in various applications, especially within 
the context of the universal relations.

\acknowledgments
GU and ZS acknowledge the Albert Einstein Center for Gravitation and Astrophysics supported 
by the Czech Science Foundation grant No. 14-37086G. MU and GT were supported by Czech 
Science Foundation grant No.17-16287S. GU, MU and ZS are grateful for very kind hospitality 
from the University of Oxford and for fruitful discussions there; the stay in Oxford was 
partly supported by grant LTC18058. They acknowledge the~internal grants of the~Silesian 
University Opava FPF SGS/14,15/2016. Some of the equations of state used in this work were 
kindly provided to the authors by J.R. Stone. The authors also acknowledge fruitful 
discussions with Marek A. Abramowicz concerning particular aspects of models of rotating 
stars and properties of their space-times. We would like to thank the referee for 
very helpful comments and suggestions and for pointing out several references that we missed 
in the previous version of our manuscript.
\software{Mathematica (version 11.3, \citet{Mathematica}) and LORENE/nrotstar (\citet{Lorene1,Lorene}}).

\bibliographystyle{aasjournal}  
\bibliography{references}   

\end{document}